\documentclass[journal,10pt,twocolumn]{IEEEtran}

\usepackage{amsmath,amssymb,bm,cite,graphicx,array}
\usepackage{placeins}
\usepackage{eso-pic}
\DeclareMathSymbol{\varDelta}{\mathalpha}{letters}{"01}
\DeclareMathSymbol{\varTheta}{\mathalpha}{letters}{"02}
\DeclareMathSymbol{\varSigma}{\mathalpha}{letters}{"06}
\newtheorem{lemma}{Lemma}
\newtheorem{proposition}{Proposition}
\newtheorem{example}{Example}

\newtheorem{definition}{Definition}

\def\IEEEproofindentspace{15.5pt}

\begin{document}
\AddToShipoutPictureFG*{%
  \AtPageUpperLeft{%
    \put(50,-20){\parbox[t]{500pt}{\normalfont\fontsize{8}{9}\selectfont
      This work has been submitted to the IEEE for possible publication. Copyright may be transferred without notice, after which this version may no longer be accessible.}}%
  }%
}

	\title{Customizing Wireless Channels via Intelligent Reflecting Surfaces: Feasibility, Fundamental Limits, and Constructive Design}
	\author{Xiyuan Liu, Qingqing Wu, Guangji Chen, Yuxuan Chen, and Wen Chen
\thanks{Xiyuan Liu, Qingqing Wu, Yuxuan Chen, and Wen Chen are with the School of Integrated Circuit, Shanghai Jiao Tong University, China (e-mails: \mbox{xiyuanliu\_sjtu@sjtu.edu.cn}; \mbox{qingqingwu@sjtu.edu.cn}; \mbox{yuxuanchen@sjtu.edu.cn}; \mbox{wenchen@sjtu.edu.cn}). Guangji Chen is with the School of Electronic and Optical Engineering, Nanjing University of Science and Technology, Nanjing 210094, China (e-mail: \mbox{guangjichen@njust.edu.cn}).}
}
	\maketitle
	\bstctlcite{channelDIY:BSTcontrol}

	\begin{abstract}
By adjusting their reflection coefficients, intelligent reflecting surfaces (IRSs) can reshape the channels between a base station (BS) and user equipment (UE). Existing studies mainly optimize IRS phase-shifts for communication objectives, leaving the physical attainability of prescribed IRS-reflected channels largely unexplored. This paper investigates channel customization with a phase-only IRS under three progressively relaxed criteria. Observing that the customization mean-squared error (MSE) has the same mathematical form as the signal MSE of a point-to-point multiple-input multiple-output system, we construct a counterpart problem to characterize individual singular-coordinate gain limits and the condition for attaining the amplitude-matching upper bound. To address multi-stream phase conflicts, we analyze principal contribution area (PCA) overlap and derive sufficient array-size conditions for disjoint PCAs under a specified focused-vector construction with parallel uniform linear arrays. We also derive necessary conditions from the two propagation links' singular values and an array-size restriction under the parallel-array Fresnel model. Both the restriction and the construction involve the BS--IRS array-size product. Numerical results show that jointly selecting right singular vectors and IRS phase-shifts reduces mean customization MSE, whereas gains exceeding propagation spectral limits cannot be realized exactly. These results provide design guidance and highlight spectral limits of phase-only IRS channel customization.
\end{abstract}


	
	\section{Introduction}
	\label{sec:introduction}
	
	Intelligent reflecting surfaces (IRSs) consist of many controllable reflecting elements, each capable of imposing an adjustable phase shift on the incident signal. By configuring these phase-shifts, IRSs can modify the cascaded channel from a base station (BS) to user equipment (UE) and thereby reconfigure the propagation environment. This capability has motivated extensive studies in wireless communications, \mbox{sensing, and networking \cite{wu2020towards,wu2021tutorial}}.
	
	Most existing studies select IRS phase-shifts for a specific communication objective. For a desired link, the phase-shifts are typically optimized to increase the received power. With undesired links or interference among UEs, they can suppress individual links \cite{jiang2022interference,li2025risaided}, reduce signal leakage \cite{santamaria2023interference}, or confine interfering signals to a common low-dimensional subspace \cite{fu2021reconfigurable,liu2024risaided,xu2023enhanced}. Supporting multiple simultaneous data streams requires considering the number of spatial subchannels and their gains, rather than only the strength of individual desired or interfering links. Related works have therefore characterized the available degrees of freedom \cite{bafghi2022degrees}, spectral efficiency and channel conditioning \cite{wang2021joint}, conditions for full spatial multiplexing \cite{zhang2022irsaided}, and spatial-multiplexing gains from multiple IRSs \cite{han2022double}. These results explain how the IRS configuration affects the number and power gains of supported spatial subchannels, but do not determine whether the entire end-to-end channel can be prescribed.\looseness=-1
	
	Several studies have more directly controlled properties related to the singular value decomposition (SVD) of the end-to-end channel. The singular values describe the gains of the spatial subchannels, whereas the right and left singular vectors specify their input and output directions, respectively. Prescribing the channel therefore requires specifying both the gains and these directions. In \cite{elmossallamy2021spatial}, the IRS phase-shifts were optimized to maximize the effective rank or minimum singular value. In \cite{chen2023channel}, the locations and phase-shifts of multiple IRSs were jointly designed to control the channel rank, truncated condition number, and singular-subchannel gains, while a sparse full-rank SVD structure was constructed in \cite{chen2023limitedfeedback} to reduce channel state information feedback. In these studies, the channel structure is obtained by optimizing a performance metric or satisfying a structural requirement, so its singular values and singular vectors are outputs of the IRS design rather than a complete set of requirements specified before the design. Thus, controlling these SVD-related properties differs from jointly prescribing the singular values and their corresponding left and right singular vectors.
	
	Whether prescribed singular subchannels can be supported depends on the sizes of the BS and IRS arrays and the propagation geometry of the BS--IRS and IRS--UE links. Effective rank measures the number of significant spatial subchannels supported by a propagation channel \cite{roy2007effective}. By modeling the transmit and receive arrays as continuous electromagnetic apertures, prior studies related the number of independent spatial subchannels to the aperture sizes and propagation distance \cite{dardari2020communicating}. Spatial separability also depends on wavefront characteristics and array geometry. Near-field studies showed that spherical wavefronts allow UEs at different angles or distances to produce distinguishable array responses \cite{lu2022near}, while studies of sparse and extremely large arrays revealed how element spacing and aperture size affect spatial resolution and channel correlation \cite{chen2024near,chen2025beamspace,chen2025effective,wang2024enhancing,wang2025analytical,wang2024extremely,sun2025differentiate}. Analysis of massive multiple-input multiple-output (MIMO) systems further unveiled that, under the corresponding channel assumptions, the normalized correlation between the channel vectors of different UEs approaches zero as the number of antennas increases \cite{marzetta2010noncooperative}. These results explained how propagation and array parameters affect the number of independent spatial subchannels and the correlation between channel vectors. However, having sufficient spatial dimensions does not by itself establish whether IRS phase control can realize prescribed channel gains and singular vectors through the given BS--IRS and IRS--UE links.\looseness=-1
	
	Existing results more directly related to channel shaping remain limited to specific design objectives or surface models. For a low-cost phase-only IRS, interference nulling has been shown to be achievable in line-of-sight (LoS) interference channels with blocked direct links \cite{jiang2022interference}. However, the available sufficient condition requires the number of IRS elements to grow exponentially with the number of UE pairs. Moreover, it guarantees only that undesired cross-links vanish, without prescribing the desired-link gains or the remaining entries of the end-to-end channel. Wavefront-control experiments have also demonstrated nearly orthogonal channels by equalizing normalized eigenvalue weights \cite{delhougne2019optimally}. This approach controls the relative distribution of singular values but not their absolute magnitudes or associated singular vectors. More general studies have analyzed arbitrary channel generation using amplitude-reconfigurable, beyond-diagonal, or fully reconfigurable surfaces \cite{alegria2024channel}, as well as the achievable singular-value regions of diagonal and beyond-diagonal architectures \cite{zhao2024mimo}. The results on arbitrary channel generation allow scattering matrices beyond the diagonal unit-modulus model of a low-cost phase-only IRS. In this model, each IRS element adjusts only the phase of the reflected signal, with no signal coupling between elements. The singular-value-region results, on the other hand, do not prescribe the associated singular vectors. Consequently, these results do not establish whether prescribed singular values and their left and right singular vectors can be jointly customized under given BS--IRS and IRS--UE links.

Motivated by this gap, this paper investigates whether, and to what extent, a prescribed channel can be customized by a low-cost phase-only IRS under these given links. The main contributions are summarized as follows:
	
	\begin{itemize}
		\item We develop an analytical framework for characterizing IRS-enabled channel customization with varying degrees of design flexibility. Specifically, we formulate three progressively relaxed customization criteria, moving beyond the restrictive requirement of exact channel matching. A key observation is that the resulting channel-customization mean-squared error (MSE) shares the same mathematical form as the signal MSE in a point-to-point MIMO system. Leveraging this structural equivalence, we construct a counterpart problem whose normalized signal MSE equals the channel-customization MSE.
		
		\item Through the counterpart problem, we characterize individual singular-coordinate gains and the phase conflicts that can arise when multiple coordinates reuse IRS elements. For the diagonal coefficient from a fixed BS-side vector to the virtual receive coordinate specified by a prescribed left singular vector, we derive the maximum magnitude over the IRS phase-shifts and the condition for attaining its amplitude-matching upper bound. For two streams whose principal contribution areas (PCAs) intersect, we identify the spatial requirement for the intersecting elements to produce nonzero desired outputs without undesired components. We also give an example in which one common phase assignment preserves the individual zero-interference outputs of both streams. The need to satisfy these phase requirements simultaneously motivates the use of disjoint PCAs. For the specified focused-vector construction with parallel uniform linear arrays (ULAs), we further derive sufficient BS- and IRS-size conditions for obtaining disjoint PCAs.
		
		\item We derive necessary conditions on the singular-value distributions of the BS--IRS and IRS--UE channels for prescribed gains and MSE tolerances. Under the parallel-array Fresnel model, these conditions yield a necessary lower bound on the BS--IRS array-size product when the allowed Frobenius-norm error is smaller than the smallest positive singular value of the required channel. Comparing this bound with the aperture requirement of the sufficient construction identifies their common dependence on the array sizes.\looseness=-1
	\end{itemize}
	
	\emph{Notation:} The symbols $[\boldsymbol x]_n$ and $\operatorname{abs}(\boldsymbol x)$ denote the $n$th entry of $\boldsymbol x$ and the vector of its element-wise magnitudes, respectively, while $\arg z$ denotes the phase of a complex scalar $z$. The norms $\|\cdot\|_{\mathrm F}$ and $\|\cdot\|_2$ denote the Frobenius norm and the Euclidean vector norm or spectral matrix norm, respectively. For a matrix $\boldsymbol A$, $\sigma_i(\boldsymbol A)$ denotes its $i$th largest singular value. Moreover, $\mathcal{CN}(\boldsymbol\mu,\boldsymbol C)$ denotes the circularly symmetric complex Gaussian distribution with mean $\boldsymbol\mu$ and covariance $\boldsymbol C$.\looseness=-1

	
	\section{IRS-Enabled Channel Customization}
	\label{sec:channel_customization_formulation}
	
	\subsection{System Model}
	
	Consider a narrowband flat-fading system in which an $M$-antenna BS communicates with $K$ single-antenna UEs with the aid of an $N$-element IRS. Let $N_{\mathrm s}$ denote the number of transmitted data streams, where $1\leq N_{\mathrm s}\leq\min\{M,K\}$. The IRS phase-shift matrix is given by $\boldsymbol\varTheta=\operatorname{diag}(e^{\jmath\theta_1},\ldots,e^{\jmath\theta_N})$, where $\theta_n\in[0,2\pi)$ denotes the phase shift of the $n$th reflecting element. Its feasible set is defined as
	\begin{equation}
		\mathcal Q_{\varTheta}\triangleq\left\{\operatorname{diag}(e^{\jmath\theta_1},\ldots,e^{\jmath\theta_N})\,\middle|\,\substack{\theta_n\in[0,2\pi),\\ n=1,\ldots,N}\right\}.
		\label{eq:irs_phase_shift_set}
	\end{equation}
	
	For any $\boldsymbol\varTheta\in\mathcal Q_{\varTheta}$, the IRS-reflected channel is expressed as
	\begin{equation}
		\boldsymbol H_{\mathrm c}(\boldsymbol\varTheta)=\boldsymbol G^{\mathrm H}\boldsymbol\varTheta\boldsymbol F\in\mathbb C^{K\times M},
		\label{eq:irs_reflected_channel}
	\end{equation}
	where $\boldsymbol F\in\mathbb C^{N\times M}$ denotes the radiative-near-field LoS BS--IRS channel, and the $k$th column of $\boldsymbol G\in\mathbb C^{N\times K}$ denotes the LoS channel from the IRS to UE $k$, modeled by either near-field spherical-wave propagation or far-field plane-wave propagation. Supporting $N_{\mathrm s}$ streams requires both $\boldsymbol F$ and $\boldsymbol G$ to have rank at least $N_{\mathrm s}$.
	
	Let $\boldsymbol W\in\mathbb C^{M\times N_{\mathrm s}}$ denote the BS precoder and $\boldsymbol s\in\mathbb C^{N_{\mathrm s}}$ denote the transmitted data vector satisfying $\mathbb E[\boldsymbol s\boldsymbol s^{\mathrm H}]=\boldsymbol I_{N_{\mathrm s}}$. The received signal vector is given by
	\begin{equation}
		\boldsymbol y=\left(\boldsymbol H_{\mathrm d}^{\mathrm H}+\boldsymbol H_{\mathrm c}(\boldsymbol\varTheta)\right)\boldsymbol W\boldsymbol s+\boldsymbol n,
		\label{eq:received_signal}
	\end{equation}
	where $\boldsymbol H_{\mathrm d}\in\mathbb C^{M\times K}$ denotes the BS--UE channel matrix and $\boldsymbol n\sim\mathcal{CN}(\boldsymbol 0,\sigma_{\mathrm n}^2\boldsymbol I_K)$ denotes the additive noise vector at the UEs, with noise variance $\sigma_{\mathrm n}^2$ at each UE. The vector $\boldsymbol y$ stacks the $K$ scalar observations; each UE accesses only its own entry, and the UEs do not jointly process $\boldsymbol y$. The direct channel $\boldsymbol H_{\mathrm d}^{\mathrm H}$ is independent of $\boldsymbol\varTheta$, whereas $\boldsymbol H_{\mathrm c}(\boldsymbol\varTheta)$ can be adjusted through the IRS phase-shifts. We therefore study $\boldsymbol H_{\mathrm c}(\boldsymbol\varTheta)$.
	
	\subsection{Customization Criteria}
	
	Let $\boldsymbol H_{\mathrm p}\in\mathbb C^{K\times M}$ denote the prescribed IRS-reflected channel. Row $k$ specifies the channel from the $M$ BS antennas to UE $k$; the rows need not be orthogonal. We assume $\operatorname{rank}(\boldsymbol H_{\mathrm p})=N_{\mathrm s}$. The SVD of $\boldsymbol H_{\mathrm p}$ is expressed as
	\begin{equation}
		\boldsymbol H_{\mathrm p}=\boldsymbol U_{\mathrm p}\boldsymbol\varSigma_{\mathrm p}\boldsymbol V_{\mathrm p}^{\mathrm H},
		\label{eq:prescribed_channel_svd}
	\end{equation}
	where $\boldsymbol U_{\mathrm p}\in\mathbb C^{K\times K}$ and $\boldsymbol V_{\mathrm p}\in\mathbb C^{M\times M}$ are unitary matrices whose first $N_{\mathrm s}$ columns contain the prescribed left and right singular vectors, respectively. Moreover, $\boldsymbol\varSigma_{\mathrm p}\in\mathbb R^{K\times M}$ is the rectangular diagonal matrix containing the $N_{\mathrm s}$ positive prescribed singular values, where $\sigma_{\mathrm p,1}\geq\cdots\geq\sigma_{\mathrm p,N_{\mathrm s}}>0$. Equivalently, the prescribed channel can be expressed as
	\begin{equation}
		\boldsymbol H_{\mathrm p}=\sum\nolimits_{\ell=1}^{N_{\mathrm s}}\sigma_{\mathrm p,\ell}\boldsymbol u_{\mathrm p,\ell}\boldsymbol v_{\mathrm p,\ell}^{\mathrm H},
		\label{eq:prescribed_channel_expansion}
	\end{equation}
	where $\boldsymbol u_{\mathrm p,\ell}$ and $\boldsymbol v_{\mathrm p,\ell}$ denote column $\ell$ of $\boldsymbol U_{\mathrm p}$ and $\boldsymbol V_{\mathrm p}$, respectively. Noncooperative UEs cannot perform a receive-side rotation requiring other UEs' signals. Moreover, independently rotating a row of $\boldsymbol H_{\mathrm p}$ changes that UE's prescribed coefficients. We therefore keep the left singular vectors fixed.

	In contrast, the BS jointly controls the signals transmitted by its antennas. Suppose that an IRS configuration realizes $\boldsymbol H_{\mathrm c}(\boldsymbol\varTheta)=\boldsymbol U_{\mathrm p}\boldsymbol\varSigma_{\mathrm p}\boldsymbol V^{\mathrm H}$ for some unitary matrix $\boldsymbol V\in\mathbb C^{M\times M}$. The BS can multiply its transmit signal vector by $\boldsymbol V\boldsymbol V_{\mathrm p}^{\mathrm H}$ before transmission. The resulting signal mapping satisfies
\begin{equation}
\boldsymbol H_{\mathrm c}(\boldsymbol\varTheta)\boldsymbol V\boldsymbol V_{\mathrm p}^{\mathrm H}
=\boldsymbol U_{\mathrm p}\boldsymbol\varSigma_{\mathrm p}\boldsymbol V^{\mathrm H}\boldsymbol V\boldsymbol V_{\mathrm p}^{\mathrm H}
\overset{(a)}{=}\boldsymbol H_{\mathrm p}.
\label{eq:bs_side_unitary_recovery}
\end{equation}
In \eqref{eq:bs_side_unitary_recovery}, (a) follows from $\boldsymbol V^{\mathrm H}\boldsymbol V=\boldsymbol I_M$ and \eqref{eq:prescribed_channel_svd}. Thus, the BS-side transformation $\boldsymbol V\boldsymbol V_{\mathrm p}^{\mathrm H}$ recovers the prescribed channel $\boldsymbol H_{\mathrm p}$.

	\begin{definition}
		Given the prescribed channel $\boldsymbol H_{\mathrm p}$, the three channel-customization criteria are defined as follows.
		\begin{enumerate}
			\item \emph{Exact channel customization}: The generated channel must exactly match the prescribed channel. Exact channel customization is achieved if there exists an IRS phase-shift matrix $\boldsymbol\varTheta\in\mathcal Q_{\varTheta}$ such that
			\begin{equation}
				\boldsymbol H_{\mathrm c}(\boldsymbol\varTheta)=\boldsymbol H_{\mathrm p}.
				\label{eq:exact_channel_customization}
			\end{equation}
			
			\item \emph{BS-side-rotated channel customization}: The generated channel must retain the prescribed singular values and left singular vectors, while the right singular vectors may be unitarily rotated. This criterion holds if there exist $\boldsymbol\varTheta\in\mathcal Q_{\varTheta}$ and a unitary matrix $\boldsymbol V\in\mathbb C^{M\times M}$ such that
			\begin{equation}
				\boldsymbol H_{\mathrm c}(\boldsymbol\varTheta)
				=\boldsymbol U_{\mathrm p}\boldsymbol\varSigma_{\mathrm p}\boldsymbol V^{\mathrm H}.
				\label{eq:isospectral_channel_customization}
			\end{equation}
			
			\item \emph{Uniformly scaled BS-side-rotated channel customization}: The generated channel may retain or reduce all prescribed singular values by a common factor $0<\kappa\leq1$. This criterion is satisfied if there exist $\boldsymbol\varTheta\in\mathcal Q_{\varTheta}$, a unitary matrix $\boldsymbol V\in\mathbb C^{M\times M}$, and $0<\kappa\leq1$ such that
			\begin{equation}
				\boldsymbol H_{\mathrm c}(\boldsymbol\varTheta)
				=\boldsymbol U_{\mathrm p}(\kappa\boldsymbol\varSigma_{\mathrm p})\boldsymbol V^{\mathrm H}.
				\label{eq:scaled_isospectral_channel_customization}
			\end{equation}
			The factor $\kappa$ preserves the prescribed singular-value ratios. A factor $\kappa>1$ increases rather than relaxes the prescribed magnitudes and is therefore not considered.
		\end{enumerate}
	\end{definition}

	\subsection{Channel-Customization Metric}
	
	To express the three criteria through a common metric, write the required channel as $\boldsymbol U_{\mathrm p}(\kappa\boldsymbol\varSigma_{\mathrm p})\boldsymbol V^{\mathrm H}$, where $0<\kappa\leq1$ and $\boldsymbol V\in\mathbb C^{M\times M}$ is unitary. Exact customization is represented by $(\kappa,\boldsymbol V)=(1,\boldsymbol V_{\mathrm p})$. Allowing $\boldsymbol V$ to be selected jointly with $\boldsymbol\varTheta$ while fixing $\kappa=1$ gives the BS-side-rotated criterion. Allowing both $\boldsymbol V$ and $\kappa\in(0,1]$ to be selected gives the uniformly scaled BS-side-rotated criterion, which includes the preceding criterion as the case $\kappa=1$. For a given $\boldsymbol\varTheta$, the channel-customization MSE, averaged over all $KM$ entries, is defined as
	\begin{equation}
		\varepsilon_{\mathrm c}(\boldsymbol\varTheta,\boldsymbol V;\kappa)\triangleq(KM)^{-1}\left\|\boldsymbol H_{\mathrm c}(\boldsymbol\varTheta)-\boldsymbol U_{\mathrm p}(\kappa\boldsymbol\varSigma_{\mathrm p})\boldsymbol V^{\mathrm H}\right\|_{\mathrm F}^2.
		\label{eq:channel_customization_mse}
	\end{equation}
	
	Using the unitary matrices $\boldsymbol U_{\mathrm p}$ and $\boldsymbol V$, the error norm in \eqref{eq:channel_customization_mse} can be equivalently transformed as
	\begin{equation}
		\begin{aligned}
			&\left\|\boldsymbol H_{\mathrm c}(\boldsymbol\varTheta)-\boldsymbol U_{\mathrm p}(\kappa\boldsymbol\varSigma_{\mathrm p})\boldsymbol V^{\mathrm H}\right\|_{\mathrm F}^2\\
			&\overset{(a)}{=}\left\|\boldsymbol U_{\mathrm p}^{\mathrm H}\left[\boldsymbol H_{\mathrm c}(\boldsymbol\varTheta)-\boldsymbol U_{\mathrm p}(\kappa\boldsymbol\varSigma_{\mathrm p})\boldsymbol V^{\mathrm H}\right]\boldsymbol V\right\|_{\mathrm F}^2\\
			&\overset{(b)}{=}\left\|\boldsymbol U_{\mathrm p}^{\mathrm H}\boldsymbol H_{\mathrm c}(\boldsymbol\varTheta)\boldsymbol V-\kappa\boldsymbol\varSigma_{\mathrm p}\right\|_{\mathrm F}^2.
		\end{aligned}
		\label{eq:unitary_invariance}
	\end{equation}
	Equality (a) follows from the unitary invariance of the Frobenius norm, and equality (b) follows from $\boldsymbol U_{\mathrm p}^{\mathrm H}\boldsymbol U_{\mathrm p}=\boldsymbol I_K$ and $\boldsymbol V^{\mathrm H}\boldsymbol V=\boldsymbol I_M$. For notational convenience, define $\boldsymbol M_{\mathrm c}(\boldsymbol\varTheta;\boldsymbol V)\triangleq\boldsymbol U_{\mathrm p}^{\mathrm H}\boldsymbol H_{\mathrm c}(\boldsymbol\varTheta)\boldsymbol V$. The channel-customization MSE can then be expressed as
	\begin{equation}
		\varepsilon_{\mathrm c}(\boldsymbol\varTheta,\boldsymbol V;\kappa)=(KM)^{-1}\left\|\boldsymbol M_{\mathrm c}(\boldsymbol\varTheta;\boldsymbol V)-\kappa\boldsymbol\varSigma_{\mathrm p}\right\|_{\mathrm F}^2.
		\label{eq:transformed_channel_customization_mse}
	\end{equation}
	
	To decompose the MSE in \eqref{eq:transformed_channel_customization_mse}, partition the unitary matrices as $\boldsymbol U_{\mathrm p}=[\boldsymbol U_{\mathrm s},\boldsymbol U_\perp]$ and $\boldsymbol V=[\boldsymbol V_{\mathrm s},\boldsymbol V_\perp]$. Here, $\boldsymbol U_{\mathrm s}$ and $\boldsymbol V_{\mathrm s}$ contain the first $N_{\mathrm s}$ columns of $\boldsymbol U_{\mathrm p}$ and $\boldsymbol V$, respectively. The matrices $\boldsymbol U_\perp$ and $\boldsymbol V_\perp$ contain the remaining $K-N_{\mathrm s}$ columns of $\boldsymbol U_{\mathrm p}$ and $M-N_{\mathrm s}$ columns of $\boldsymbol V$, respectively. The columns of $\boldsymbol U_\perp$ and $\boldsymbol V_\perp$ form orthonormal bases for $\operatorname{span}(\boldsymbol U_{\mathrm s})^\perp$ and $\operatorname{span}(\boldsymbol V_{\mathrm s})^\perp$, respectively. To identify individual entries of $\boldsymbol M_{\mathrm c}$, let $\boldsymbol u_i$ and $\boldsymbol v_j$ denote columns $i$ and $j$ of $\boldsymbol U_{\mathrm p}$ and $\boldsymbol V$, respectively. Accordingly, $\boldsymbol M_{\mathrm c}$ is partitioned as
	\begin{equation}
		\boldsymbol M_{\mathrm c}=\boldsymbol U_{\mathrm p}^{\mathrm H}\boldsymbol H_{\mathrm c}(\boldsymbol\varTheta)\boldsymbol V
		\triangleq\begin{bmatrix}\boldsymbol M_{\mathrm s}&\boldsymbol M_{\mathrm s\perp}\\\boldsymbol M_{\perp\mathrm s}&\boldsymbol M_{\perp\perp}\end{bmatrix}.
		\label{eq:modal_coupling_partition}
	\end{equation}
	The four blocks satisfy $\boldsymbol M_{\mathrm s}\in\mathbb C^{N_{\mathrm s}\times N_{\mathrm s}}$, $\boldsymbol M_{\mathrm s\perp}\in\mathbb C^{N_{\mathrm s}\times(M-N_{\mathrm s})}$, $\boldsymbol M_{\perp\mathrm s}\in\mathbb C^{(K-N_{\mathrm s})\times N_{\mathrm s}}$, and $\boldsymbol M_{\perp\perp}\in\mathbb C^{(K-N_{\mathrm s})\times(M-N_{\mathrm s})}$. Denote the $(i,\ell)$th entry of $\boldsymbol M_{\mathrm s}$ by $m_{i,\ell}$. Since the only nonzero entries of $\boldsymbol\varSigma_{\mathrm p}$ are $\sigma_{\mathrm p,1},\ldots,\sigma_{\mathrm p,N_{\mathrm s}}$ on its principal diagonal, \eqref{eq:transformed_channel_customization_mse} can be decomposed as
	\begin{equation}
		\begin{aligned}
			KM\varepsilon_{\mathrm c}={}&\sum\nolimits_{\ell=1}^{N_{\mathrm s}}\left|m_{\ell,\ell}-\kappa\sigma_{\mathrm p,\ell}\right|^2+\sum\nolimits_{\substack{i,\ell=1\\i\neq\ell}}^{N_{\mathrm s}}|m_{i,\ell}|^2\\
			&+\left\|\boldsymbol M_{\perp\mathrm s}\right\|_{\mathrm F}^2+\left\|\boldsymbol M_{\mathrm s\perp}\right\|_{\mathrm F}^2+\left\|\boldsymbol M_{\perp\perp}\right\|_{\mathrm F}^2.
		\end{aligned}
		\label{eq:channel_customization_mse_decomposition}
	\end{equation}
	
	A zero MSE in \eqref{eq:transformed_channel_customization_mse} indicates that the generated IRS-reflected channel satisfies the adopted channel-customization criterion without error. When strictly zero MSE cannot be attained or is unnecessary for an engineering implementation, the adopted criterion is regarded as satisfied within a prescribed tolerance if
	\begin{equation}
		\varepsilon_{\mathrm c}(\boldsymbol\varTheta,\boldsymbol V;\kappa)\leq\tau_{\mathrm c},
		\label{eq:channel_customization_tolerance}
	\end{equation}
	where $\tau_{\mathrm c}\geq0$ denotes the channel-customization tolerance.

\subsection{Counterpart Problem}

The original customization problem is to find an admissible design $(\boldsymbol\varTheta,\boldsymbol V,\kappa)$ satisfying the MSE requirement in \eqref{eq:channel_customization_tolerance}. Equation~\eqref{eq:transformed_channel_customization_mse} compares $\boldsymbol M_{\mathrm c}=\boldsymbol U_{\mathrm p}^{\mathrm H}\boldsymbol H_{\mathrm c}(\boldsymbol\varTheta)\boldsymbol V$ with $\kappa\boldsymbol\varSigma_{\mathrm p}$.

To interpret this matrix comparison in terms of signal transmission, introduce an auxiliary random input vector $\boldsymbol s_{\mathrm{cp}}\in\mathbb C^M$ satisfying $\mathbb E[\boldsymbol s_{\mathrm{cp}}\boldsymbol s_{\mathrm{cp}}^{\mathrm H}]=\boldsymbol I_M$. The product $\boldsymbol M_{\mathrm c}\boldsymbol s_{\mathrm{cp}}$ describes transmission of $\boldsymbol V\boldsymbol s_{\mathrm{cp}}$ through $\boldsymbol H_{\mathrm c}(\boldsymbol\varTheta)$, followed by receive combining with $\boldsymbol U_{\mathrm p}^{\mathrm H}$. The output is
\begin{equation}
\boldsymbol y_{\mathrm c}
=\boldsymbol U_{\mathrm p}^{\mathrm H}\boldsymbol G^{\mathrm H}
\boldsymbol\varTheta\boldsymbol F\boldsymbol V\boldsymbol s_{\mathrm{cp}}
=\boldsymbol M_{\mathrm c}(\boldsymbol\varTheta;\boldsymbol V)\boldsymbol s_{\mathrm{cp}}.
\label{eq:counterpart_signal_model}
\end{equation}
The desired output corresponding to $\kappa\boldsymbol\varSigma_{\mathrm p}$ is
\begin{equation}
\boldsymbol y_{\mathrm p}
=\kappa\boldsymbol\varSigma_{\mathrm p}\boldsymbol s_{\mathrm{cp}}.
\label{eq:counterpart_desired_signal}
\end{equation}
Noise is omitted because \eqref{eq:transformed_channel_customization_mse} measures customization error.

Since $\mathbb E[\boldsymbol s_{\mathrm{cp}}\boldsymbol s_{\mathrm{cp}}^{\mathrm H}]=\boldsymbol I_M$, the expected squared signal error equals the squared Frobenius norm of the difference $\boldsymbol M_{\mathrm c}-\kappa\boldsymbol\varSigma_{\mathrm p}$. After division by $KM$, as in \eqref{eq:channel_customization_mse}, we obtain
\begin{equation}
\begin{aligned}
&\frac{1}{KM}\mathbb E\!\left[
\|\boldsymbol y_{\mathrm c}-\boldsymbol y_{\mathrm p}\|_2^2
\right]
=\frac{1}{KM}\mathbb E\!\left[
\|(\boldsymbol M_{\mathrm c}-\kappa\boldsymbol\varSigma_{\mathrm p})
\boldsymbol s_{\mathrm{cp}}\|_2^2
\right]\\
&\quad=\frac{1}{KM}
\|\boldsymbol M_{\mathrm c}-\kappa\boldsymbol\varSigma_{\mathrm p}\|_{\mathrm F}^2
\overset{(a)}{=}\varepsilon_{\mathrm c}(\boldsymbol\varTheta,\boldsymbol V;\kappa).
\end{aligned}
\label{eq:counterpart_mse}
\end{equation}
In \eqref{eq:counterpart_mse}, (a) follows from \eqref{eq:transformed_channel_customization_mse}. We define the counterpart problem as choosing the same admissible $(\boldsymbol\varTheta,\boldsymbol V,\kappa)$ to make this normalized signal MSE at most $\tau_{\mathrm c}$. By \eqref{eq:counterpart_mse}, a design satisfies this requirement if and only if it satisfies the original customization requirement in \eqref{eq:channel_customization_tolerance}.

In the counterpart problem, the $K$ UEs' received signals are jointly processed by the fixed matrix $\boldsymbol U_{\mathrm p}^{\mathrm H}$ from the SVD of $\boldsymbol H_{\mathrm p}$. This joint processing allows the $K$ UEs to be viewed as a single receiver, which forms a point-to-point MIMO system with the $M$-antenna BS. Joint processing is introduced only in the counterpart problem; the UEs in the original customization problem remain noncooperative.


	\section{IRS Phase Control for Channel Customization}
\label{sec:prescribed_singular_subchannel_feasibility}

The MSE decomposition in \eqref{eq:channel_customization_mse_decomposition} expresses channel customization in terms of the desired subchannel coefficients and undesired received components. All these terms depend on the same IRS phase-shifts, so a phase-shift assignment that increases one desired coefficient can also affect the other streams. To examine this coupling, we first characterize the maximum gain of one subchannel from its received contributions through individual IRS elements. Using a common amplitude threshold, we define principal contribution areas to identify elements whose individual contribution magnitudes reach that threshold. When two such areas overlap, the phase-shifts on their shared elements must serve both subchannels. We examine when these elements can generate both desired outputs without undesired received components. To avoid imposing competing phase-shift assignments on shared elements, we are thus led to select the matrix of right singular vectors $\boldsymbol V$ so that these areas are disjoint, while the receive combining vectors remain prescribed.

\subsection{Customization of a Single Subchannel}
\label{subsec:integrated_III_A_HEAD}

For the required channel matrix $\boldsymbol U_{\mathrm p}(\kappa\boldsymbol\varSigma_{\mathrm p})\boldsymbol V^{\mathrm H}$ in \eqref{eq:channel_customization_mse}, $\boldsymbol v_\ell$ is its $\ell$th right singular vector. In the counterpart problem, it serves as the BS transmit vector, while receive combining uses the prescribed vector $\boldsymbol u_\ell$. For a unit symbol transmitted along $\boldsymbol v_\ell$, the required received vector before combining is $\kappa\sigma_{\mathrm p,\ell}\boldsymbol u_\ell$. Thus, the prescribed singular value $\kappa\sigma_{\mathrm p,\ell}>0$ is the desired scalar gain along $\boldsymbol u_\ell$ in the counterpart problem. Let $\boldsymbol x_\ell=\boldsymbol F\boldsymbol v_\ell$ and $\boldsymbol\alpha_\ell=\boldsymbol G\boldsymbol u_\ell$. The corresponding received coefficient is
\begin{equation}
m_{\ell,\ell}(\boldsymbol\varTheta)
=\boldsymbol\alpha_\ell^{\mathrm H}\boldsymbol\varTheta\boldsymbol x_\ell
=\sum\nolimits_{n=1}^{N}e^{\jmath\theta_n}\alpha_{\ell,n}^{*}x_{\ell,n}.
\label{eq:integrated_individual}
\end{equation}
For a unit symbol on stream $\ell$, the summand $e^{\jmath\theta_n}\alpha_{\ell,n}^{*}x_{\ell,n}$ is its received coefficient along $\boldsymbol u_\ell$ through IRS element $n$. Its magnitude $|\alpha_{\ell,n}^{*}x_{\ell,n}|$ depends on the channel matrices $\boldsymbol F$ and $\boldsymbol G$ and the selected right singular vector and prescribed receive vector. The IRS changes only its phase through $\theta_n$.

Approximate customization permits nonzero MSE under \eqref{eq:channel_customization_tolerance}. We identify the elements with contribution magnitudes at least a common threshold so that their phase alignment and the error from the remaining reflections can be analyzed separately. For an amplitude threshold $\gamma>0$, common to all $N_{\mathrm s}$ subchannels, define the PCA of subchannel $\ell$ as
\begin{equation}
\mathcal S_\ell=\left\{n\in\{1,\ldots,N\}:|\alpha_{\ell,n}^{*}x_{\ell,n}|\ge \gamma\right\}.
\label{eq:integrated_pca}
\end{equation}
For fixed channel matrices and fixed right singular and receive vectors, the IRS phase-shift multiplies the contribution through element $n$ by $e^{\jmath\theta_n}$. It changes the phase of this contribution but not its magnitude, since $|e^{\jmath\theta_n}\alpha_{\ell,n}^{*}x_{\ell,n}|=|\alpha_{\ell,n}^{*}x_{\ell,n}|$. Membership in $\mathcal S_\ell$ is determined by comparing this individual magnitude with $\gamma$, so changing the IRS phase-shifts alone cannot move an element into or out of $\mathcal S_\ell$. The sum in \eqref{eq:integrated_individual} can nevertheless change because its terms add with different phases. The intersection $\mathcal S_i\cap\mathcal S_\ell$ identifies elements whose received amplitudes reach $\gamma$ along both prescribed directions. All terms in \eqref{eq:integrated_individual}, including those below $\gamma$, remain in \mbox{the complete MSE.}\looseness=-1

For given $\boldsymbol v_\ell$ and $\boldsymbol u_\ell$, maximizing $|m_{\ell,\ell}|$ requires all nonzero terms in \eqref{eq:integrated_individual} to have the same phase. The maximum is
\begin{equation}
\max_{\boldsymbol\varTheta\in\mathcal Q_{\varTheta}}|m_{\ell,\ell}(\boldsymbol\varTheta)|
=\sum\nolimits_{n=1}^{N}|\alpha_{\ell,n}^{*}x_{\ell,n}|
\overset{(a)}{\le}\|\boldsymbol\alpha_\ell\|_2\|\boldsymbol x_\ell\|_2.
\label{eq:integrated_maximum}
\end{equation}
In \eqref{eq:integrated_maximum}, (a) follows from the Cauchy--Schwarz inequality. The sum is attained by choosing
\begin{equation}
\theta_n=\bigl(\arg\alpha_{\ell,n}-\arg x_{\ell,n}\bigr)\bmod 2\pi
\label{eq:integrated_alignment}
\end{equation}
for every $n$ with $\alpha_{\ell,n}^{*}x_{\ell,n}\ne0$. Substitution into \eqref{eq:integrated_individual} gives $e^{\jmath\theta_n}\alpha_{\ell,n}^{*}x_{\ell,n}=|\alpha_{\ell,n}^{*}x_{\ell,n}|$, so reflected signals add coherently. Phases of zero-valued terms do not affect $m_{\ell,\ell}$. For these fixed vectors, zero-error matching of the single diagonal coefficient requires $\kappa\sigma_{\mathrm p,\ell}\leq\sum_{n=1}^{N}|\alpha_{\ell,n}^{*}x_{\ell,n}|$. For a nonzero tolerance, \eqref{eq:channel_customization_mse_decomposition} gives $|m_{\ell,\ell}-\kappa\sigma_{\mathrm p,\ell}|\leq\sqrt{KM\tau_{\mathrm c}}$ whenever $\varepsilon_{\mathrm c}\leq\tau_{\mathrm c}$. Combining this inequality with \eqref{eq:integrated_maximum} yields the single-coefficient amplitude requirement $\kappa\sigma_{\mathrm p,\ell}\leq\sum_{n=1}^{N}|\alpha_{\ell,n}^{*}x_{\ell,n}|+\sqrt{KM\tau_{\mathrm c}}$. This requirement concerns only $m_{\ell,\ell}$ for the specified $\boldsymbol v_\ell$ and $\boldsymbol u_\ell$; it does not control the other terms in the customization MSE. Section~\ref{sec:channel_orthogonality} instead derives necessary conditions on the singular-value distributions of $\boldsymbol F$ and $\boldsymbol G$ that apply regardless of the selected right singular vectors and IRS phase-shifts.

The sum in \eqref{eq:integrated_maximum} is the inner product of the magnitude vectors $\operatorname{abs}(\boldsymbol x_\ell)$ and $\operatorname{abs}(\boldsymbol\alpha_\ell)$. For fixed nonzero norms, the upper bound $\|\boldsymbol\alpha_\ell\|_2\|\boldsymbol x_\ell\|_2$ is attained if and only if
\begin{equation}
\frac{\operatorname{abs}(\boldsymbol x_\ell)}{\|\boldsymbol x_\ell\|_2}
=\frac{\operatorname{abs}(\boldsymbol\alpha_\ell)}{\|\boldsymbol\alpha_\ell\|_2}.
\label{eq:integrated_magnitudes}
\end{equation}
Under BS-side rotation, a vector $\boldsymbol v_\ell$ satisfying \eqref{eq:integrated_magnitudes}, together with the IRS phase-shifts in \eqref{eq:integrated_alignment}, attains the upper bound $\|\boldsymbol\alpha_\ell\|_2\|\boldsymbol x_\ell\|_2$. Equation~\eqref{eq:integrated_magnitudes} is therefore a condition for attaining the norm-product upper bound, rather than a requirement for matching every prescribed singular value. If the prescribed value $\kappa\sigma_{\mathrm p,\ell}$ lies below this upper bound, matching $m_{\ell,\ell}=\kappa\sigma_{\mathrm p,\ell}$ may still be possible when \eqref{eq:integrated_magnitudes} does not hold. Here, $m_{\ell,\ell}$ is the received coefficient along the prescribed vector $\boldsymbol u_\ell$ for input $\boldsymbol v_\ell$; it is not in general a singular value of \mbox{the generated channel $\boldsymbol H_{\mathrm c}$.}\looseness=-1

\begin{figure}[!t]
\centering
\includegraphics[width=\columnwidth]{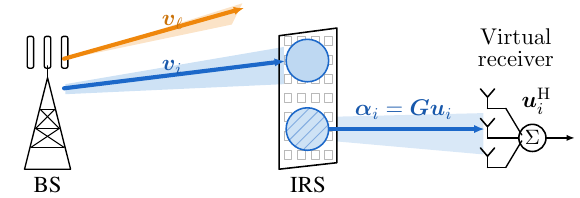}
\caption{Zero subchannel gain under every IRS phase configuration.}
\label{fig:individual_singular_subchannel_mismatch}
\end{figure}

Fig.~\ref{fig:individual_singular_subchannel_mismatch} shows two cases with zero subchannel gain for every IRS phase-shift assignment. In the first case, the orange direction $\boldsymbol v_\ell$ misses the IRS, giving $\boldsymbol x_\ell=\boldsymbol0$. Every term in \eqref{eq:integrated_individual} is then zero because no signal is incident on the IRS. In the second case, the blue direction $\boldsymbol v_i$ illuminates the upper blue region, where $\boldsymbol x_i=\boldsymbol F\boldsymbol v_i$ has nonzero entries. However, $\boldsymbol\alpha_i=\boldsymbol G\boldsymbol u_i$ has nonzero entries only in the lower striped region. These two sets of elements are disjoint, so $\alpha_{i,n}^{*}x_{i,n}=0$ for every $n$ despite the nonzero incident signal. The diagonal IRS phase-shift matrix cannot transfer the signal from the illuminated elements to those contributing to the prescribed receive direction. Both cases use idealized zeros; any nonzero sidelobe terms of a finite array remain in \eqref{eq:integrated_maximum}.

\subsection{Simultaneous Customization of Multiple Subchannels}
\label{subsec:integrated_III_B_HEAD}

The phase alignment in \eqref{eq:integrated_alignment} maximizes the gain of one subchannel. If $\mathcal S_i\cap\mathcal S_\ell=\varnothing$ for all $i\ne\ell$, the contributions within each PCA can be phase-aligned for its associated subchannel using a separate set of phase-shift variables. When $\mathcal C_{i,\ell}=\mathcal S_i\cap\mathcal S_\ell\ne\varnothing$, the phase-shifts on the intersecting elements must serve both streams simultaneously. In particular, $e^{\jmath\theta_n}$ multiplies both desired subchannel coefficients through element $n$, $\alpha_{i,n}^{*}x_{i,n}$ and $\alpha_{\ell,n}^{*}x_{\ell,n}$, as well as the cross-stream coefficients through that element, $\alpha_{i,n}^{*}x_{\ell,n}$ and $\alpha_{\ell,n}^{*}x_{i,n}$. These shared phase-shifts couple the two designs. We examine when they support both desired gains without undesired components.

For a unit symbol transmitted on stream $j\in\{i,\ell\}$, let $\boldsymbol\psi_j^{\mathcal C}\in\mathbb C^K$ denote the received signal vector generated by the elements in $\mathcal C_{i,\ell}$. It is given by
\begin{equation}
\boldsymbol\psi_j^{\mathcal C}(\boldsymbol\varTheta)
=\sum\nolimits_{n\in\mathcal C_{i,\ell}}e^{\jmath\theta_n}x_{j,n}
\boldsymbol G^{\mathrm H}\boldsymbol e_n,
\quad j\in\{i,\ell\}.
\label{eq:integrated_regional_output}
\end{equation}
Here, $\boldsymbol e_n$ is column $n$ of $\boldsymbol I_N$, and $\boldsymbol G^{\mathrm H}\boldsymbol e_n$ is the IRS--UE channel vector associated with element $n$. The projection onto $\boldsymbol u_j$ gives the desired received coefficient $\boldsymbol u_j^{\mathrm H}\boldsymbol\psi_j^{\mathcal C}$. The orthogonal projection contains all undesired receive components. For unit-power symbols on the two streams, the total power in these undesired components is
\begin{equation}
P_{i,\ell}^{\mathcal C}
=\sum\nolimits_{j\in\{i,\ell\}}
\left\|(\boldsymbol I_K-\boldsymbol u_j\boldsymbol u_j^{\mathrm H})
\boldsymbol\psi_j^{\mathcal C}\right\|_2^2.
\label{eq:integrated_regional_undesired}
\end{equation}
The projector $\boldsymbol I_K-\boldsymbol u_j\boldsymbol u_j^{\mathrm H}$ removes the component along $\boldsymbol u_j$. Hence, nonzero desired gains with $P_{i,\ell}^{\mathcal C}=0$ require $\boldsymbol\psi_i^{\mathcal C}$ and $\boldsymbol\psi_\ell^{\mathcal C}$ to be nonzero multiples of $\boldsymbol u_i$ and $\boldsymbol u_\ell$, respectively. Both received vectors belong to the span of $\{\boldsymbol G^{\mathrm H}\boldsymbol e_n:n\in\mathcal C_{i,\ell}\}$ by \eqref{eq:integrated_regional_output}. This span must therefore contain both prescribed receive vectors and have dimension at least two. This condition alone does not ensure that the fixed incident amplitudes and one IRS phase-shift matrix can produce both vectors.

To compare the alignment phase-shifts \eqref{eq:integrated_alignment} on $\mathcal C_{i,\ell}$, define
\begin{equation}
[\boldsymbol z_j^{\mathcal C}]_n
=\frac{\alpha_{j,n}^{*}x_{j,n}}{|\alpha_{j,n}^{*}x_{j,n}|},
\quad n\in\mathcal C_{i,\ell}.
\label{eq:integrated_regional_phase}
\end{equation}
The following example uses one IRS phase assignment on $\mathcal C_{i,\ell}$ to maximize both $|\boldsymbol u_i^{\mathrm H}\boldsymbol\psi_i^{\mathcal C}|$ and $|\boldsymbol u_\ell^{\mathrm H}\boldsymbol\psi_\ell^{\mathcal C}|$, while giving $P_{i,\ell}^{\mathcal C}=0$.

\begin{example}
Suppose that, for $j\in\{i,\ell\}$, applying stream $j$'s alignment phase-shifts on $\mathcal C_{i,\ell}$ produces only its prescribed receive direction, i.e.,
\begin{equation}
\sum\nolimits_{n\in\mathcal C_{i,\ell}}[\boldsymbol z_j^{\mathcal C}]_n^{*}
x_{j,n}\boldsymbol G^{\mathrm H}\boldsymbol e_n
=\boldsymbol u_j\sum\nolimits_{n\in\mathcal C_{i,\ell}}
|\alpha_{j,n}^{*}x_{j,n}|.
\label{eq:integrated_regional_premise}
\end{equation}
Both streams can retain these individually aligned gains under one phase assignment if
\begin{equation}
\boldsymbol z_i^{\mathcal C}
=e^{\jmath\delta}\boldsymbol z_\ell^{\mathcal C},
\qquad \delta\in[0,2\pi).
\label{eq:integrated_regional_relation}
\end{equation}
For each $n\in\mathcal C_{i,\ell}$, choose $e^{\jmath\theta_n}=[\boldsymbol z_i^{\mathcal C}]_n^{*}$. The choice aligns the received coefficients of stream $i$ along $\boldsymbol u_i$, and \eqref{eq:integrated_regional_relation} makes it $e^{-\jmath\delta}$ times the alignment for stream $\ell$, giving
\begin{gather}
\boldsymbol\psi_i^{\mathcal C}=\boldsymbol u_i\sum\nolimits_{n\in\mathcal C_{i,\ell}}|\alpha_{i,n}^{*}x_{i,n}|,
\label{eq:integrated_regional_result}\\
\boldsymbol\psi_\ell^{\mathcal C}=e^{-\jmath\delta}\boldsymbol u_\ell\sum\nolimits_{n\in\mathcal C_{i,\ell}}|\alpha_{\ell,n}^{*}x_{\ell,n}|.
\label{eq:integrated_regional_result_ell}
\end{gather}
Thus, $P_{i,\ell}^{\mathcal C}=0$, and both desired magnitudes are preserved. Under the criteria allowing BS-side rotation, the phase factor of stream $\ell$ can be removed by replacing $\boldsymbol v_\ell$ with $e^{\jmath\delta}\boldsymbol v_\ell$. This rotates the entire stream, including signals through elements outside $\mathcal C_{i,\ell}$, which remain in \mbox{the complete MSE.} Exact customization fixes $\boldsymbol V_{\mathrm p}$ and does not permit this BS-side rotation.

\end{example}

Equation~\eqref{eq:integrated_regional_relation} requires the relative phase between $\alpha_{i,n}^{*}x_{i,n}$ and $\alpha_{\ell,n}^{*}x_{\ell,n}$ to be constant over $n\in\mathcal C_{i,\ell}$, so both individually maximizing assignments coincide up to a stream phase. Varying relative phases can still allow simultaneous gain enhancement at lower amplitudes. Thus, Example 1 does not establish PCA separation as necessary for customization.

To avoid assigning competing phase alignments on shared PCA elements, we seek orthonormal right singular vectors with $\mathcal S_i\cap\mathcal S_\ell=\varnothing$ for all $i\ne\ell$. Each $\mathcal S_\ell$ then has its own phase-shift variables for aligning the desired coefficient. The complete MSE still includes received signals through elements below the threshold, components outside each prescribed receive direction, and outputs associated with $\boldsymbol V_\perp$ whose prescribed values are zero, as specified in \eqref{eq:channel_customization_mse_decomposition}.

\section{Sufficient Channel Customization Conditions}
\label{sec:simultaneous_singular_subchannel_customization}

Disjoint PCAs permit separate phase-shift assignments for the desired subchannel coefficients, but the remaining reflections still contribute to the customization MSE. We therefore select the required channel's right singular vectors through the counterpart problem and seek sufficient array-size conditions that bound this complete error.

\subsection{Channel Customization With Disjoint PCAs}

We consider the criteria allowing $\boldsymbol V$ to be selected with prescribed $\boldsymbol U_{\mathrm p}$ and $\boldsymbol\varSigma_{\mathrm p}$. The BS-side-rotated criterion fixes $\kappa=1$. For the uniformly scaled criterion, choose $\kappa\in(0,1]$ before the construction. Orthogonality alone does not ensure disjoint PCAs, since both $|\alpha_{i,n}^{*}x_{i,n}|$ and $|\alpha_{\ell,n}^{*}x_{\ell,n}|$ may reach $\gamma$ at one element even when $\boldsymbol v_i^{\mathrm H}\boldsymbol v_\ell=0$. We focus the incident signals at separated locations and assign regions around their intensity peaks. Separate focal-spacing and width parameters permit expanding these regions without changing the transmit vectors. Bounding the remaining contributions, together with common-phase matching and nulling, gives a sufficient test for meeting the full-channel MSE tolerance.

We first derive the BS--IRS channel matrix to calculate the incident amplitudes produced by the focused vectors. Consider parallel BS and IRS ULAs facing each other. The line joining their centers is perpendicular to both arrays and has length $r_{\mathrm{BI}}$. Let $d_{\mathrm B}$ and $d_{\mathrm I}$ denote the BS and IRS element spacings, respectively. The coordinates along each array, measured from its center, are
\begin{gather}
p_m=\left(m-\frac{M+1}{2}\right)d_{\mathrm B},\quad m=1,\ldots,M,
\label{eq:integrated_coordinates}\\
q_n=\left(n-\frac{N+1}{2}\right)d_{\mathrm I},\quad n=1,\ldots,N.
\label{eq:integrated_irs_coordinates}
\end{gather}
The distance from BS antenna $m$ to IRS element $n$ is $\sqrt{r_{\mathrm{BI}}^2+(q_n-p_m)^2}$. We use its second-order expansion to retain the curvature of the propagation phase. Let $a=((M-1)d_{\mathrm B}+(N-1)d_{\mathrm I})/2$ be the largest transverse separation between the two arrays' elements. With wavelength $\lambda$, this Fresnel approximation requires $a/r_{\mathrm{BI}}\ll1$ and the phase-error bound $\pi a^4/(4\lambda r_{\mathrm{BI}}^3)$ to lie within the chosen modeling accuracy. Approximating the path amplitudes by a \mbox{common coefficient $\beta\ne0$ gives}
\begin{equation}
[\boldsymbol F]_{n,m}
=\beta\exp\left[-\jmath\frac{2\pi r_{\mathrm{BI}}}{\lambda}
-\jmath\frac{\pi(q_n-p_m)^2}{\lambda r_{\mathrm{BI}}}\right].
\label{eq:integrated_channel}
\end{equation}
The sufficient conditions use \eqref{eq:integrated_channel}, with either a near-field or far-field IRS--UE model.

For parallel arrays whose center-to-center line is oblique, let $\vartheta_{\mathrm{BI}}$ be its signed angle from the common broadside direction in the plane containing the arrays, measured toward increasing array coordinates, with $|\vartheta_{\mathrm{BI}}|<\pi/2$. The second-order expansion of the propagation distance is $r_{\mathrm{BI}}+(q_n-p_m)\sin\vartheta_{\mathrm{BI}}+(q_n-p_m)^2\cos^2\vartheta_{\mathrm{BI}}/(2r_{\mathrm{BI}})$. The linear term gives separate phase factors for BS antennas and IRS elements, which can be incorporated into $\boldsymbol V$ and the IRS phase-shifts. The quadratic term then has the form in \eqref{eq:integrated_channel} with projected spacings $d_{\mathrm B}\cos\vartheta_{\mathrm{BI}}$ and $d_{\mathrm I}\cos\vartheta_{\mathrm{BI}}$. The distance-approximation error must be checked for this oblique geometry; the fourth-order bound above applies to the facing arrays. Nonparallel arrays require their own distance expansion.

For the facing arrays, let $\rho_\ell$ denote subchannel $\ell$'s focal coordinate measured from the IRS center. Its angle from BS broadside is $\arctan(\rho_\ell/r_{\mathrm{BI}})$, allowing off-axis focusing. Compensate the quadratic phase from each BS antenna to this point by choosing
\begin{equation}
[\boldsymbol v_\ell]_m
=M^{-1/2}
\exp\left[\jmath\frac{\pi(\rho_\ell-p_m)^2}{\lambda r_{\mathrm{BI}}}\right].
\label{eq:integrated_focus}
\end{equation}
Each entry has magnitude $M^{-1/2}$, giving unit norm. Substitution into $\boldsymbol x_\ell=\boldsymbol F\boldsymbol v_\ell$ yields the phase difference $(\rho_\ell-p_m)^2-(q_n-p_m)^2=\rho_\ell^2-q_n^2+2p_m(q_n-\rho_\ell)$. Cancellation of the quadratic terms in $p_m$ leaves a geometric sum with phase increment $2\pi d_{\mathrm B}(q_n-\rho_\ell)/(\lambda r_{\mathrm{BI}})$. Define $\varDelta_{\mathrm{BI}}=\lambda r_{\mathrm{BI}}/(Md_{\mathrm B})$ and $\nu_{\ell,n}=(q_n-\rho_\ell)/\varDelta_{\mathrm{BI}}$. The sum gives
\begin{equation}
\begin{aligned}
|x_{\ell,n}|^2
&=\frac{|\beta|^2}{M}
\left|\sum\nolimits_{m=1}^{M}
e^{\jmath 2\pi p_m(q_n-\rho_\ell)/(\lambda r_{\mathrm{BI}})}
\right|^2\\
&=|\beta|^2M
\left|\frac{\sin(\pi\nu_{\ell,n})}
{M\sin(\pi\nu_{\ell,n}/M)}\right|^2 .
\end{aligned}
\label{eq:integrated_field_power}
\end{equation}
Evaluate the ratio by continuity at removable singularities. For $M\ge2$, its first zeros are at $\nu=\pm1$, so $\varDelta_{\mathrm{BI}}$ is the distance from the focus to either first null and the null-to-null width is $2\varDelta_{\mathrm{BI}}$. A null need not coincide with an IRS element. This geometric series follows from the Fresnel phase cancellation and applies to near-field focusing.

The vectors in \eqref{eq:integrated_focus} must be mutually orthogonal because they are columns of the unitary right singular matrix $\boldsymbol V$. Their inner product is
\begin{equation}
\boldsymbol v_i^{\mathrm H}\boldsymbol v_\ell
=\frac{e^{\jmath\pi(\rho_\ell^2-\rho_i^2)/(\lambda r_{\mathrm{BI}})}}{M}
\sum\limits_{m=1}^{M}e^{-\jmath2\pi p_m(\rho_\ell-\rho_i)/(\lambda r_{\mathrm{BI}})}.
\label{eq:integrated_focus_inner}
\end{equation}
The same geometric-series calculation gives zero when $\rho_\ell-\rho_i=s\varDelta_{\mathrm{BI}}$, where $s$ is an integer satisfying $0<|s|<M$. We therefore choose equally spaced focal points as
\begin{equation}
\rho_\ell=\left(\ell-\frac{N_{\mathrm s}+1}{2}\right)
J\varDelta_{\mathrm{BI}},\qquad \ell=1,\ldots,N_{\mathrm s},
\label{eq:integrated_focus_positions}
\end{equation}
where $J$ is a positive integer satisfying $N_{\mathrm s}J\le M$. For distinct streams, $s=(\ell-i)J$ then obeys $0<|s|<M$, ensuring orthogonality. The selected vectors form $\boldsymbol V_{\mathrm s}$ and can be completed to a unitary matrix $\boldsymbol V=[\boldsymbol V_{\mathrm s},\boldsymbol V_\perp]$. The remaining columns have zero prescribed gains and remain in the MSE.

The intensity in \eqref{eq:integrated_field_power} repeats with spatial period $M\varDelta_{\mathrm{BI}}$. In this construction, we include the neighborhoods of repeated peaks lying on the IRS. Define the normalized distance to the nearest peak by $t_{\ell,n}=\min_{k\in\mathbb Z}|\nu_{\ell,n}-kM|$, so $0\le t_{\ell,n}\le M/2$. For a positive integer $Q\le J$, assign
\begin{equation}
\mathcal I_\ell(Q)
=\{n\in\{1,\ldots,N\}:t_{\ell,n}<Q/2\}.
\label{eq:integrated_interval}
\end{equation}
The windows have width $Q\varDelta_{\mathrm{BI}}$ and centers $\rho_\ell+kM\varDelta_{\mathrm{BI}}$. Neighboring streams' peaks are separated by $J\varDelta_{\mathrm{BI}}$ within a period and by $[M-(N_{\mathrm s}-1)J]\varDelta_{\mathrm{BI}}\ge J\varDelta_{\mathrm{BI}}$ across its boundary. Hence, $Q\le J$ makes the assigned sets disjoint over all periods. These sets are determined by position, whereas the PCAs are amplitude-defined. \mbox{A later bound ensures $\mathcal S_\ell\subseteq\mathcal I_\ell(Q)$.}\looseness=-1

\begin{figure}[!t]
\centering
\includegraphics[width=\columnwidth]{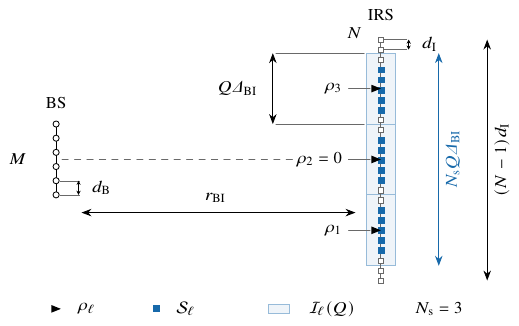}
\caption{IRS element allocation for three subchannels.}
\label{fig:irs_pca_packing_condition}
\end{figure}

To accommodate the windows around the central focal points, it suffices that
\begin{equation}
[(N_{\mathrm s}-1)J+Q]\varDelta_{\mathrm{BI}}
\le (N-1)d_{\mathrm I}.
\label{eq:integrated_aperture_range}
\end{equation}
No upper aperture restriction is imposed to exclude repeated peaks, since their neighborhoods are included through \eqref{eq:integrated_interval}. The Fresnel modeling conditions stated above must still hold for the physical arrays. Fig.~\ref{fig:irs_pca_packing_condition} illustrates the central windows for three subchannels with $Q=J$. Black triangles mark focal points, bounded segments mark assigned windows, and blue squares mark PCA elements. Other peak neighborhoods, if present on the IRS, are assigned by the same rule.

\subsection{Customization Error and Assigned Region Width}

To relate the window width $Q\varDelta_{\mathrm{BI}}$ to the customization MSE, split each column of $\boldsymbol F\boldsymbol V_{\mathrm s}$ into components inside and outside its assigned region as follows.
\begin{gather}
[\boldsymbol X_{\mathrm{in}}]_{n,\ell}=\begin{cases}x_{\ell,n},&n\in\mathcal I_\ell(Q),\\0,&n\notin\mathcal I_\ell(Q),\end{cases}
\label{eq:integrated_field_split}\\
\boldsymbol X_{\mathrm{out}}=\boldsymbol F\boldsymbol V_{\mathrm s}-\boldsymbol X_{\mathrm{in}}.
\label{eq:integrated_field_out}
\end{gather}
The matrix $\boldsymbol X_{\mathrm{in}}\in\mathbb C^{N\times N_{\mathrm s}}$ retains the assigned entries, while $\boldsymbol X_{\mathrm{out}}$ contains all remaining incident components. Their total squared magnitude is
\begin{equation}
E_{\mathrm{out}}(Q)
=\|\boldsymbol X_{\mathrm{out}}\|_{\mathrm F}^2
=\sum\nolimits_{\ell=1}^{N_{\mathrm s}}
\sum\nolimits_{n\notin\mathcal I_\ell(Q)}|x_{\ell,n}|^2.
\label{eq:integrated_energy_out}
\end{equation}
This algebraic split does not change the IRS operation, in which each element reflects all incident signals with one phase shift. Thus, $\boldsymbol X_{\mathrm{out}}$ includes signals on other subchannels' regions and outside all assigned regions.

To relate $E_{\mathrm{out}}(Q)$ to the customization MSE, right-multiply the error matrix in \eqref{eq:channel_customization_mse} by $\boldsymbol V$. This operation preserves its Frobenius norm because $\boldsymbol V$ is unitary. Let $\boldsymbol\varSigma_{\mathrm{p,s}}\in\mathbb R^{N_{\mathrm s}\times N_{\mathrm s}}$ contain the positive prescribed singular values on its diagonal. Separating the columns associated with $\boldsymbol V_{\mathrm s}$ and $\boldsymbol V_\perp$ \mbox{and substituting \eqref{eq:integrated_field_split}--\eqref{eq:integrated_field_out} gives}
\begin{equation}
\begin{aligned}
KM\varepsilon_{\mathrm c}
={}&\left\|\boldsymbol G^{\mathrm H}\boldsymbol\varTheta\boldsymbol X_{\mathrm{in}}
-\boldsymbol U_{\mathrm s}(\kappa\boldsymbol\varSigma_{\mathrm{p,s}})
+\boldsymbol G^{\mathrm H}\boldsymbol\varTheta\boldsymbol X_{\mathrm{out}}
\right\|_{\mathrm F}^2\\
&+\left\|\boldsymbol G^{\mathrm H}\boldsymbol\varTheta
\boldsymbol F\boldsymbol V_\perp\right\|_{\mathrm F}^2 .
\end{aligned}
\label{eq:integrated_complete_error}
\end{equation}
The first norm contains the assigned-component mismatch and remaining contributions; the second measures output along right singular directions with zero prescribed gains. Both retain all $K$ receive components.

We next identify conditions under which the components outside the assigned regions are the only remaining source of customization error. Suppose that the same $\boldsymbol\varTheta\in\mathcal Q_{\varTheta}$ satisfies
\begin{gather}
\boldsymbol G^{\mathrm H}\boldsymbol\varTheta\boldsymbol X_{\mathrm{in}}=\boldsymbol U_{\mathrm s}(\kappa\boldsymbol\varSigma_{\mathrm{p,s}}),
\label{eq:integrated_joint_conditions}\\
\boldsymbol G^{\mathrm H}\boldsymbol\varTheta\boldsymbol F\boldsymbol V_\perp=\boldsymbol0.
\label{eq:integrated_null_condition}
\end{gather}
Equation~\eqref{eq:integrated_joint_conditions} jointly matches the prescribed gains and left singular directions. Equation~\eqref{eq:integrated_null_condition} nulls output on the complementary right singular directions, allowing nonzero $\boldsymbol\varTheta\boldsymbol F\boldsymbol V_\perp$ in $\ker(\boldsymbol G^{\mathrm H})$; it is absent when $N_{\mathrm s}=M$. \mbox{Substituting both into \eqref{eq:integrated_complete_error} yields}\looseness=-1
\begin{equation}
\varepsilon_{\mathrm c}
=\frac{\|\boldsymbol G^{\mathrm H}\boldsymbol\varTheta
\boldsymbol X_{\mathrm{out}}\|_{\mathrm F}^2}{KM}
\le\frac{\|\boldsymbol G\|_2^2E_{\mathrm{out}}(Q)}{KM}.
\label{eq:integrated_mse_tail}
\end{equation}
The bound uses $\|\boldsymbol A\boldsymbol B\|_{\mathrm F}\le\|\boldsymbol A\|_2\|\boldsymbol B\|_{\mathrm F}$ together with $\|\boldsymbol G^{\mathrm H}\|_2=\|\boldsymbol G\|_2$ and $\|\boldsymbol\varTheta\|_2=1$. It holds for any relative phases in $\boldsymbol X_{\mathrm{out}}$. The common-phase conditions \eqref{eq:integrated_joint_conditions}--\eqref{eq:integrated_null_condition} do not follow from \mbox{PCA separation alone.}\looseness=-1

We next bound $E_{\mathrm{out}}(Q)$ without restricting the IRS to one intensity period. Choose an integer $k_{\ell,n}$ nearest to $\nu_{\ell,n}/M$, resolving ties toward the integer of smaller magnitude. The samples assigned to the same $k$ form a cell around one peak. Let $D_\ell$ be the number of these cells containing IRS elements for stream $\ell$, and let $D=\sum_{\ell=1}^{N_{\mathrm s}}D_\ell$. These finite counts depend on the array geometry and focal spacing, but not on $Q$. Periodicity permits replacing $|\nu_{\ell,n}|$ by $t_{\ell,n}$ in the field magnitude. Since $\sin u\ge2u/\pi$ for $0\le u\le\pi/2$, \eqref{eq:integrated_field_power} gives
\begin{equation}
|x_{\ell,n}|^2
\le\frac{|\beta|^2M}{4t_{\ell,n}^2},
\qquad 0<t_{\ell,n}\le M/2.
\label{eq:integrated_point_bound}
\end{equation}
Outside $\mathcal I_\ell(Q)$, $t_{\ell,n}\ge Q/2>0$. On either side of a peak within one cell, successive distances are separated by $d_{\mathrm I}/\varDelta_{\mathrm{BI}}$. If that side contains an element outside the assigned window, let $t_+\ge Q/2$ be its nearest distance. The decreasing \mbox{function $t^{-2}$ then yields}
\begin{equation}
\begin{aligned}
\sum_{\substack{n\text{ on one side}\\n\notin\mathcal I_\ell(Q)}}\frac{1}{t_{\ell,n}^{2}}
&\le\frac{1}{t_+^2}+
\frac{\varDelta_{\mathrm{BI}}}{d_{\mathrm I}}
\int_{t_+}^{\infty}\frac{dt}{t^2}\\
&\le\frac{4}{Q^2}+\frac{2\varDelta_{\mathrm{BI}}}{d_{\mathrm I}Q}.
\end{aligned}
\label{eq:integrated_tail_sum}
\end{equation}
The first term is retained separately; each later term is bounded by the integral over the preceding grid interval divided by its width. Empty sides contribute zero. Summing both sides in each occupied cell and then all streams gives
\begin{equation}
E_{\mathrm{out}}(Q)
\le|\beta|^2MD
\left(\frac{2}{Q^2}+\frac{\varDelta_{\mathrm{BI}}}{d_{\mathrm I}Q}\right).
\label{eq:integrated_energy_bound}
\end{equation}
The infinite integral only bounds a finite sum within each cell; \eqref{eq:integrated_point_bound} is used solely for distances no greater than $M/2$. With the arrays and focal spacing fixed, increasing $Q$ enlarges nested assigned sets while leaving $\boldsymbol V_{\mathrm s}$ unchanged. Therefore, the actual $E_{\mathrm{out}}(Q)$, as well as this bound, cannot increase. Increasing $N$ changes the sampling grid and may change $D$ and $\boldsymbol G$, so the error must be evaluated again at the new size.

The same pointwise estimate ensures PCA containment. Let $\alpha_{\max}=\max_{\ell,n}|\alpha_{\ell,n}|$. Outside $\mathcal I_\ell(Q)$, $t_{\ell,n}\ge Q/2$, so \eqref{eq:integrated_point_bound} gives $|\alpha_{\ell,n}^{*}x_{\ell,n}|\le\alpha_{\max}|\beta|\sqrt M/Q$. Therefore,
\begin{equation}
\frac{\alpha_{\max}|\beta|\sqrt M}{Q}<\gamma
\quad\Longrightarrow\quad
\mathcal S_\ell\subseteq\mathcal I_\ell(Q).
\label{eq:integrated_pca_containment}
\end{equation}
Since the assigned regions are disjoint, this inclusion ensures $\mathcal S_i\cap\mathcal S_\ell=\varnothing$ for $i\ne\ell$. The threshold $\gamma$ is used only to identify the PCAs; no term is discarded from the MSE bound \eqref{eq:integrated_mse_tail}. Combining that bound with \eqref{eq:integrated_energy_bound} gives the following \mbox{sufficient error condition.}

\begin{lemma}\label{lem:finite_array_disjoint_pca}
Consider \eqref{eq:integrated_channel} with $1\le N_{\mathrm s}\le\min\{K,N\}$, $M\ge\max\{2,N_{\mathrm s}\}$, prescribed $\boldsymbol U_{\mathrm s}$ and $\boldsymbol\varSigma_{\mathrm{p,s}}$, and a specified admissible $\kappa$. Select the vectors \eqref{eq:integrated_focus} at \eqref{eq:integrated_focus_positions}, with positive integers $J,Q$ satisfying $Q\le J$ and $N_{\mathrm s}J\le M$. Suppose one $\boldsymbol\varTheta\in\mathcal Q_{\varTheta}$ satisfies both \eqref{eq:integrated_joint_conditions} and \eqref{eq:integrated_null_condition}. For $\tau_{\mathrm c}>0$, if
\begin{equation}
\frac{|\beta|^2D\|\boldsymbol G\|_2^2}{K}
\left(\frac{2}{Q^2}+\frac{\lambda r_{\mathrm{BI}}}
{Md_{\mathrm B}d_{\mathrm I}Q}\right)
\le\tau_{\mathrm c},
\label{eq:integrated_sufficient_error}
\end{equation}
then the resulting channel satisfies $\varepsilon_{\mathrm c}\le\tau_{\mathrm c}$. If the strict inequality in \eqref{eq:integrated_pca_containment} also holds, its PCAs are disjoint.
\end{lemma}

\begin{IEEEproof}
The condition $N_{\mathrm s}J\le M$ ensures orthonormal selected vectors, so completing them to $\boldsymbol V$ permits \eqref{eq:integrated_complete_error}. The assumed common phase-shift matrix matches the assigned components and nulls the complementary directions. Substituting \eqref{eq:integrated_energy_bound} into \eqref{eq:integrated_mse_tail} then gives
\begin{equation}
\varepsilon_{\mathrm c}\overset{(a)}{\le}\frac{|\beta|^2D\|\boldsymbol G\|_2^2}{K}
\left(\frac{2}{Q^2}+\frac{\varDelta_{\mathrm{BI}}}{d_{\mathrm I}Q}\right)
\overset{(b)}{\le}\tau_{\mathrm c}.
\label{eq:integrated_sufficient_proof}
\end{equation}
Here, (a) uses \eqref{eq:integrated_energy_bound}, and (b) uses \eqref{eq:integrated_sufficient_error}. Finally, \eqref{eq:integrated_pca_containment} places the PCAs in the disjoint assigned sets.
\end{IEEEproof}

\subsection{Sufficient Array Size Conditions}

For given arrays and a focal spacing $J$ with $N_{\mathrm s}J\le M$, the count $D$ is known. Using $2/Q^2\le2/Q$ for $Q\ge1$, an integer width sufficient for \eqref{eq:integrated_sufficient_error} satisfies
\begin{equation}
Q\ge
\max\left\{1,\left\lceil
\frac{|\beta|^2D\|\boldsymbol G\|_2^2}{K\tau_{\mathrm c}}
\left(2+\frac{\lambda r_{\mathrm{BI}}}{Md_{\mathrm B}d_{\mathrm I}}\right)
\right\rceil\right\}.
\label{eq:integrated_q_choice}
\end{equation}
The width can be increased up to $J$ without changing the focal vectors or overlapping the assigned regions. The following proposition combines this error requirement with the aperture needed to contain the central windows.

\begin{proposition}\label{prop:finite_array_disjoint_pca}
Consider \eqref{eq:integrated_channel} with $1\le N_{\mathrm s}\le\min\{K,N\}$, $M\ge\max\{2,N_{\mathrm s}\}$, prescribed $\boldsymbol U_{\mathrm s}$ and $\boldsymbol\varSigma_{\mathrm{p,s}}$, a specified admissible $\kappa$, and $\tau_{\mathrm c}>0$. Choose positive integers $J,Q$ satisfying $N_{\mathrm s}J\le M$, $Q\le J$, and \eqref{eq:integrated_q_choice}. Select the right singular vectors \eqref{eq:integrated_focus} at \eqref{eq:integrated_focus_positions}. If one $\boldsymbol\varTheta\in\mathcal Q_{\varTheta}$ satisfies \eqref{eq:integrated_joint_conditions} and \eqref{eq:integrated_null_condition}, and \mbox{the IRS size satisfies}
\begin{equation}
N\ge1+\left\lceil\frac{[(N_{\mathrm s}-1)J+Q]\lambda r_{\mathrm{BI}}}{Md_{\mathrm B}d_{\mathrm I}}\right\rceil,
\label{eq:integrated_n_range}
\end{equation}
then the central assigned windows fit within the IRS and the resulting channel satisfies $\varepsilon_{\mathrm c}\le\tau_{\mathrm c}$. If the strict inequality in \eqref{eq:integrated_pca_containment} also holds, its PCAs are disjoint.
\end{proposition}
\begin{IEEEproof}
Substituting $\varDelta_{\mathrm{BI}}=\lambda r_{\mathrm{BI}}/(Md_{\mathrm B})$ into \eqref{eq:integrated_aperture_range} and using integer $N$ gives \eqref{eq:integrated_n_range}. Since $Q\ge1$, the left-hand side of \eqref{eq:integrated_sufficient_error} is at most $|\beta|^2D\|\boldsymbol G\|_2^2(2+\varDelta_{\mathrm{BI}}/d_{\mathrm I})/(KQ)$, which does not exceed $\tau_{\mathrm c}$ by \eqref{eq:integrated_q_choice}. The remaining hypotheses are those of Lemma~\ref{lem:finite_array_disjoint_pca}, proving the result.
\end{IEEEproof}

Assigning all peak neighborhoods removes the upper bound on $N$ needed by a single-period tail estimate. Evaluate $D$ and $\boldsymbol G$ at the tested size and geometry, and check the common-phase conditions for the chosen width. If no $Q\le J$ meets \eqref{eq:integrated_q_choice}, inserting the finite sum \eqref{eq:integrated_energy_out} into \eqref{eq:integrated_mse_tail} gives a less conservative test. Failure of either test does not prove infeasibility. Enlarging an assigned region changes the algebraic split, not \mbox{the amplitude-defined PCA.}\looseness=-1

The aperture requirement also shows how the two array sizes enter this construction. Rearranging \eqref{eq:integrated_aperture_range} gives
\begin{equation}
M(N-1)\ge
\frac{[(N_{\mathrm s}-1)J+Q]\lambda r_{\mathrm{BI}}}{d_{\mathrm B}d_{\mathrm I}}.
\label{eq:integrated_product_general}
\end{equation}
In the special case $Q=J$, with $d_{\mathrm B}=\lambda/2$ and $d_{\mathrm I}=\lambda/4$ \cite{liu2026irsspacing}, this reduces to
\begin{equation}
M(N-1)\ge\frac{8N_{\mathrm s}Qr_{\mathrm{BI}}}{\lambda}.
\label{eq:integrated_product_quarter}
\end{equation}
For fixed stream count, window parameters, distance, wavelength, and spacings, increasing $M$ narrows the windows through $\varDelta_{\mathrm{BI}}\propto1/M$, while increasing $N$ accommodates them. Thus, the aperture condition allows a tradeoff through $M(N-1)$, approximately $MN$ for $N\gg1$. This product does not by itself fix the error bound or the common-phase feasibility, which also depend on $D$, $\boldsymbol G$, and the prescribed channel. Nor does this construction determine the minimum sizes over all possible customization designs.

\section{Necessary Channel Customization Conditions}
\label{sec:channel_orthogonality}

The sufficient conditions may require larger arrays than the original problem needs. When violated, necessary conditions exclude customization regardless of the design, and comparing the two helps assess the sufficient construction's conservativeness. We also seek to determine whether the size-product dependence in its aperture condition follows from propagation or only from the focused-vector design. We therefore derive spectral necessary conditions and relate them to array sizes under the Fresnel model. This comparison helps distinguish the array-size requirements imposed by propagation from those introduced by the sufficient construction.

\subsection{Singular-Value Conditions on $\boldsymbol F$ and $\boldsymbol G$}

Consider $M,N,K\ge N_{\mathrm s}\ge2$ and a specified $0<\kappa\le1$. The target in \eqref{eq:channel_customization_mse} has smallest positive singular value $\kappa\sigma_{\mathrm p,N_{\mathrm s}}$. By the SVD approximation theorem, deleting a singular component with this value gives the minimum squared Frobenius error, $\kappa^2\sigma_{\mathrm p,N_{\mathrm s}}^2$, over matrices of rank at most $N_{\mathrm s}-1$. Thus, to require the generated channel to retain at least $N_{\mathrm s}$ nonzero singular values, we restrict the MSE tolerance to
\begin{equation}
0\le\tau_{\mathrm c}
<
\frac{\kappa^2\sigma_{\mathrm p,N_{\mathrm s}}^2}{KM}.
\label{eq:necessary_1}
\end{equation}
Under \eqref{eq:necessary_1}, no channel of rank below $N_{\mathrm s}$ can meet \eqref{eq:channel_customization_tolerance}. At or above this threshold, the MSE criterion permits a target approximation of rank at most $N_{\mathrm s}-1$, although that approximation need not be realizable by the IRS.

For a design meeting \eqref{eq:channel_customization_tolerance}, the Frobenius norm of the channel error is at most $\sqrt{KM\tau_{\mathrm c}}$. Each singular-value perturbation is bounded by the spectral norm of this error, which is no larger than its Frobenius norm. Hence, $\left|\sigma_{N_{\mathrm s}}(\boldsymbol H_{\mathrm c})-\kappa\sigma_{\mathrm p,N_{\mathrm s}}\right|\le\sqrt{KM\tau_{\mathrm c}}$, and \eqref{eq:necessary_1} gives a positive lower bound on $\sigma_{N_{\mathrm s}}(\boldsymbol H_{\mathrm c})$. The matrix-product bounds in \cite[Corollary~3.3]{zhao2024mimo} then yield the necessary inequalities
\begin{gather}
\kappa\sigma_{\mathrm p,N_{\mathrm s}}-\sqrt{KM\tau_{\mathrm c}}\le\sigma_{N_{\mathrm s}}(\boldsymbol H_{\mathrm c})\le\sigma_1(\boldsymbol G)\sigma_{N_{\mathrm s}}(\boldsymbol F),
\label{eq:necessary_2}\\
\kappa\sigma_{\mathrm p,N_{\mathrm s}}-\sqrt{KM\tau_{\mathrm c}}\le\sigma_{N_{\mathrm s}}(\boldsymbol H_{\mathrm c})\le\sigma_1(\boldsymbol F)\sigma_{N_{\mathrm s}}(\boldsymbol G).
\label{eq:necessary_2_G}
\end{gather}

To verify that the upper bound in \eqref{eq:necessary_2} applies to the diagonal IRS in \eqref{eq:irs_reflected_channel}, retain the $N_{\mathrm s}-1$ largest singular components of $\boldsymbol F$. Let $\boldsymbol u_{\mathrm F,i}$ and $\boldsymbol v_{\mathrm F,i}$ denote its left and right singular vectors, respectively. The truncation and the rank bound on its product with $\boldsymbol G^{\mathrm H}\boldsymbol\varTheta$ are given by
\begin{gather}
\boldsymbol F_{N_{\mathrm s}-1}=\sum\nolimits_{i=1}^{N_{\mathrm s}-1}\sigma_i(\boldsymbol F)\boldsymbol u_{\mathrm F,i}\boldsymbol v_{\mathrm F,i}^{\mathrm H},
\label{eq:necessary_3}\\
\operatorname{rank}\!\left(\boldsymbol G^{\mathrm H}\boldsymbol\varTheta\boldsymbol F_{N_{\mathrm s}-1}\right)\le N_{\mathrm s}-1.
\label{eq:necessary_truncated_rank}
\end{gather}
The rank bound follows from the fact that matrix multiplication cannot increase the rank of $\boldsymbol F_{N_{\mathrm s}-1}$.

The SVD approximation theorem gives $\sigma_{N_{\mathrm s}}(\boldsymbol H_{\mathrm c})$ as the minimum spectral-norm error when approximating $\boldsymbol H_{\mathrm c}$ by a matrix of rank at most $N_{\mathrm s}-1$. The matrix in \eqref{eq:necessary_truncated_rank} is one such approximation. Its error therefore gives the \mbox{following upper bound.}
\begin{equation}
\begin{aligned}
&\sigma_{N_{\mathrm s}}(\boldsymbol H_{\mathrm c})\overset{(a)}{\le}\sigma_1\!\left(\boldsymbol H_{\mathrm c}-\boldsymbol G^{\mathrm H}\boldsymbol\varTheta\boldsymbol F_{N_{\mathrm s}-1}\right)\\
&\quad=\sigma_1\!\left(\boldsymbol G^{\mathrm H}\boldsymbol\varTheta(\boldsymbol F-\boldsymbol F_{N_{\mathrm s}-1})\right)\\
&\quad\overset{(b)}{\le}\sigma_1(\boldsymbol G^{\mathrm H}\boldsymbol\varTheta)\sigma_1(\boldsymbol F-\boldsymbol F_{N_{\mathrm s}-1})\overset{(c)}{=}\sigma_1(\boldsymbol G)\sigma_{N_{\mathrm s}}(\boldsymbol F).
\end{aligned}
\label{eq:necessary_4}
\end{equation}
Here, (a) applies the minimum approximation error, and (b) uses submultiplicativity of the spectral norm. In (c), right multiplication by the unitary matrix $\boldsymbol\varTheta$ preserves the singular values of $\boldsymbol G^{\mathrm H}$, while the largest singular value of the discarded part $\boldsymbol F-\boldsymbol F_{N_{\mathrm s}-1}$ is $\sigma_{N_{\mathrm s}}(\boldsymbol F)$. Applying the same argument to $\boldsymbol H_{\mathrm c}^{\mathrm H}=\boldsymbol F^{\mathrm H}\boldsymbol\varTheta^{\mathrm H}\boldsymbol G$, with $\boldsymbol G$ truncated instead, proves \eqref{eq:necessary_2_G}. The two bounds hold for every admissible diagonal IRS matrix without further assumptions on $\boldsymbol F$ or $\boldsymbol G$.

Equations \eqref{eq:necessary_2}--\eqref{eq:necessary_2_G} require at least $N_{\mathrm s}$ nonzero singular values in both propagation matrices, with minimum magnitudes determined by the prescribed gain and tolerance. IRS phase adjustment cannot compensate for a violation.

To relate the requirement on $\boldsymbol F$ to the array sizes, we use the ratio $\|\boldsymbol F\|_{\mathrm F}^2/\sigma_1^2(\boldsymbol F)$, for which an upper bound can be obtained from the array geometry. The required singular values impose a lower bound on this ratio. Since $\sigma_2(\boldsymbol F),\ldots,\sigma_{N_{\mathrm s}}(\boldsymbol F)$ are all at least $\sigma_{N_{\mathrm s}}(\boldsymbol F)$, we have
\begin{equation}
\begin{aligned}
\frac{\|\boldsymbol F\|_{\mathrm F}^2}{\sigma_1^2(\boldsymbol F)}
&\overset{(a)}{=}
1+\sum\nolimits_{i=2}^{\min\{M,N\}}
\frac{\sigma_i^2(\boldsymbol F)}{\sigma_1^2(\boldsymbol F)}\\
&\overset{(b)}{\ge}
1+(N_{\mathrm s}-1)
\frac{\left(\kappa\sigma_{\mathrm p,N_{\mathrm s}}
-\sqrt{KM\tau_{\mathrm c}}\right)^2}
{\sigma_1^2(\boldsymbol F)\sigma_1^2(\boldsymbol G)} .
\end{aligned}
\label{eq:necessary_5}
\end{equation}
Equality (a) expresses the squared Frobenius norm as the sum of squared singular values. Inequality (b) retains the $N_{\mathrm s}-1$ terms with $i=2,\ldots,N_{\mathrm s}$ and applies \eqref{eq:necessary_2} to each. 

For the model in \eqref{eq:integrated_channel}, set $c_{\mathrm{BI}}=MN d_{\mathrm B}d_{\mathrm I}/(\lambda r_{\mathrm{BI}})$. Combining the upper bound on this ratio derived in Appendix~\ref{app:necessary_array_bound} with \eqref{eq:necessary_5} gives the following \mbox{necessary condition on $MN$.}

\begin{proposition}\label{prop:cascaded_spectral_upper_bound} Consider the BS and IRS arrays in \eqref{eq:integrated_coordinates} and \eqref{eq:integrated_irs_coordinates}, with the channel matrix $\boldsymbol F$ given by \eqref{eq:integrated_channel}, $M,N,K\ge N_{\mathrm s}\ge2$, and $\sigma_1(\boldsymbol F)\sigma_1(\boldsymbol G)>0$. For a specified $\kappa$ and a tolerance satisfying \eqref{eq:necessary_1}, any design meeting \eqref{eq:channel_customization_tolerance} must satisfy
\begin{equation}
MN\ge
\frac{(N_{\mathrm s}-1)\lambda r_{\mathrm{BI}}}
{8d_{\mathrm B}d_{\mathrm I}}
\frac{
\left(\kappa\sigma_{\mathrm p,N_{\mathrm s}}
-\sqrt{KM\tau_{\mathrm c}}\right)^2
}{
\sigma_1^2(\boldsymbol F)\sigma_1^2(\boldsymbol G)
}.
\label{eq:necessary_6}
\end{equation}

\end{proposition}

\begin{IEEEproof} See Appendix~\ref{app:necessary_array_bound}. \end{IEEEproof}

The appendix bounds $\|\boldsymbol F\|_{\mathrm F}^2/\sigma_1^2(\boldsymbol F)$ above by $1+8c_{\mathrm{BI}}$. Comparing this geometric bound with \eqref{eq:necessary_5} and subtracting $1$ gives \eqref{eq:necessary_6}, where $N_{\mathrm s}-1$ counts the required singular values other than the largest. No right singular vectors or IRS phase-shifts are prescribed in this argument.

This geometric dependence is consistent with the spatial-resolution limitation of active BS arrays separating LoS UEs \cite{marzetta2010noncooperative}. Here, BS--IRS propagation involves two arrays, so both $M$ and $N$ enter the bound. The IRS--UE channel enters \eqref{eq:necessary_6} through $\sigma_1(\boldsymbol G)$ and \eqref{eq:necessary_2_G} through $\sigma_{N_{\mathrm s}}(\boldsymbol G)$.

\subsection{Implications for Channel Customization}

The array-size requirement in Proposition~\ref{prop:cascaded_spectral_upper_bound} and the spectral bounds \eqref{eq:necessary_2}--\eqref{eq:necessary_2_G} are necessary, but do not ensure that one admissible IRS phase-shift matrix meets all prescribed gains and left singular directions within the MSE tolerance. We compare them with Proposition~\ref{prop:finite_array_disjoint_pca}.

For this comparison, keep $\boldsymbol U_{\mathrm s}$, the prescribed singular values, $\kappa$, $\tau_{\mathrm c}>0$, $\boldsymbol F$, and $\boldsymbol G$ fixed, and allow the right singular vectors to be selected. The sufficient construction uses \eqref{eq:integrated_focus}--\eqref{eq:integrated_focus_positions}, with $Q\le J$ and $N_{\mathrm s}J\le M$, and requires the aperture inequality \eqref{eq:integrated_aperture_range}, common-phase conditions \eqref{eq:integrated_joint_conditions}--\eqref{eq:integrated_null_condition}, and error bound \eqref{eq:integrated_sufficient_error}. PCA separation additionally follows from the strict inequality in \eqref{eq:integrated_pca_containment}; it is not assumed in \mbox{the necessary condition.}\looseness=-1

We now compare the necessary bound \eqref{eq:necessary_6} in Proposition~\ref{prop:cascaded_spectral_upper_bound} with the aperture requirement \eqref{eq:integrated_product_general} used in Proposition~\ref{prop:finite_array_disjoint_pca}. The latter is expressed in terms of $M(N-1)$. Multiplying it by $N/(N-1)$, with $N\ge2$, gives the comparison
\begin{gather}
MN\ge\frac{(N_{\mathrm s}-1)\lambda r_{\mathrm{BI}}}{8d_{\mathrm B}d_{\mathrm I}}\frac{\left(\kappa\sigma_{\mathrm p,N_{\mathrm s}}-\sqrt{KM\tau_{\mathrm c}}\right)^2}{\sigma_1^2(\boldsymbol F)\sigma_1^2(\boldsymbol G)},
\label{eq:necessary_14}\\
MN\ge\frac{N}{N-1}\frac{[(N_{\mathrm s}-1)J+Q]\lambda r_{\mathrm{BI}}}{d_{\mathrm B}d_{\mathrm I}}.
\label{eq:necessary_constructive_product}
\end{gather}

Both bounds contain $\lambda r_{\mathrm{BI}}/(d_{\mathrm B}d_{\mathrm I})$, but their other factors differ; the shared dependence does not establish that the bounds are close. Changing $M$ or $N$ changes $\boldsymbol F$, while $N$ also changes $\boldsymbol G$. For positive tolerance, $\sqrt{KM\tau_{\mathrm c}}$ varies with $M$. The spectra and parameters $J,Q,D$ must therefore be evaluated for each pair $(M,N)$. These size inequalities alone neither determine minimum array sizes nor quantify the gap to the sufficient conditions.

For the spacings $d_{\mathrm B}=\lambda/2$ and $d_{\mathrm I}=\lambda/4$ used in \eqref{eq:integrated_product_quarter}, substituting into \eqref{eq:necessary_6} gives
\begin{equation}
MN\ge
\frac{(N_{\mathrm s}-1)r_{\mathrm{BI}}}{\lambda}
\frac{\left(\kappa\sigma_{\mathrm p,N_{\mathrm s}}
-\sqrt{KM\tau_{\mathrm c}}\right)^2}
{\sigma_1^2(\boldsymbol F)\sigma_1^2(\boldsymbol G)} .
\label{eq:necessary_15}
\end{equation}
This size requirement is conditional on the specified target scaling $\kappa$. With $\tau_{\mathrm c}=0$ and fixed prescribed singular values, $\boldsymbol F$, and $\boldsymbol G$, the right-hand side is proportional to $\kappa^2$ and tends to zero as $\kappa\to0$. For fixed positive $\tau_{\mathrm c}$, taking $\kappa\to0$ instead violates \eqref{eq:necessary_1}. The necessary size requirement must therefore be interpreted together with the target scaling and MSE tolerance.

\section{Numerical Results}
	\label{sec:numerical_results}

Numerical results illustrate PCA separation, channel customization, and the BS--IRS singular-value distribution. Unless otherwise stated, $\lambda=0.03\,\mathrm{m}$, $d_{\mathrm B}=\lambda/2$, and $d_{\mathrm I}=\lambda/4$. IRS phase-shifts are continuous with unit-magnitude reflection coefficients. Power ratios are reported in decibels (dB). The normalized customization MSE is $KM\varepsilon_{\mathrm c}(\boldsymbol\varTheta,\boldsymbol V;1)/\|\boldsymbol H_{\mathrm p}\|_{\mathrm F}^{2}$, with $\varepsilon_{\mathrm c}$ defined in \eqref{eq:channel_customization_mse}.

\subsection{Channel Customization Error}

\begin{figure}[!t]
		\centering
		\includegraphics[width=\columnwidth,trim=0 11.47pt 0 9.97pt,clip]{\detokenize{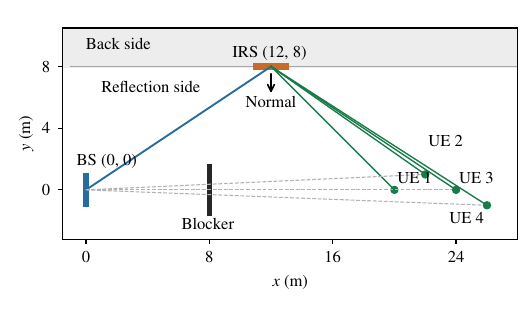}}
		\caption{BS, IRS, and UE locations in the customization experiment.}
		\label{fig:numerical_geometry}
	\end{figure}

Fig.~\ref{fig:numerical_geometry} uses $M=128$, $N=256$, and $K=4$, with BS and IRS centers at $(0,0)$ m and $(12,8)$ m. The UEs are at $(20,0)$ m, $(22,1)$ m, $(24,0)$ m, and $(26,-1)$ m. A blocker at $x=8$ m over $-1.5\leq y\leq1.5$ m blocks the direct BS--UE paths but leaves both reflected hops unobstructed; diffraction is neglected. Both $\boldsymbol F$ and $\boldsymbol G$ use spherical-wave LoS propagation with distance attenuation. For each of 300 targets, the four left singular vectors form a random orthonormal basis in the column space of $\boldsymbol G^{\mathrm H}$. The four right singular vectors form a random orthonormal set in the span of the right singular vectors of $\boldsymbol F$ whose singular values satisfy $\sigma_i(\boldsymbol F)/\sigma_1(\boldsymbol F)\geq0.1$. The four singular values are drawn independently from the uniform distribution on $[0.4,1]$, then rescaled to the \mbox{prescribed target energy.}

The fixed-right-vector scheme minimizes $\|\boldsymbol H_{\mathrm c}(\boldsymbol\varTheta)-\boldsymbol U_{\mathrm p}\boldsymbol\varSigma_{\mathrm p}\boldsymbol V_{\mathrm p}^{\mathrm H}\|_{\mathrm F}^2$ by updating one IRS phase-shift at a time with the others fixed, using eight initializations and at most 80 sweeps. The joint scheme alternates the same phase update with an SVD-based unitary Procrustes update of the right singular vectors. It uses six initializations, including the fixed-vector solution, and at most 40 outer iterations with five phase sweeps each. For numerical scaling, both generated and prescribed matrices are divided by the same product of two path-amplitude factors, evaluated at the BS--IRS center-to-center distance and the IRS-center--UE-centroid distance, respectively. Each scheme retains its lowest-error run and stops when the decrease in squared Frobenius error is below $10^{-9}$ times the larger of the preceding objective and unity.

\begin{figure}[!t]
		\centering
		\includegraphics[width=\columnwidth,trim=0 3.00pt 0 12.00pt,clip]{\detokenize{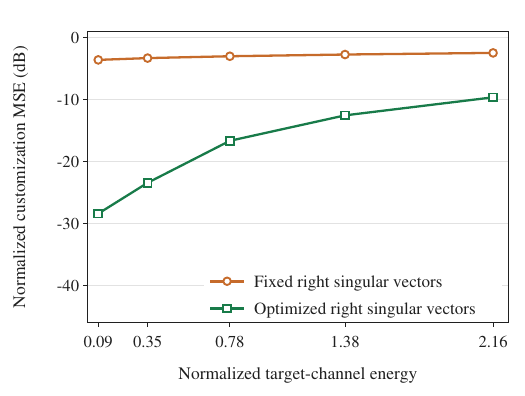}}
		\caption{Normalized customization MSE versus target-channel energy.}
		\label{fig:numerical_customization_mse}
	\end{figure}

Fig.~\ref{fig:numerical_customization_mse} compares two schemes for customizing channels with the same prescribed $\boldsymbol U_{\mathrm p}$ and $\boldsymbol\varSigma_{\mathrm p}$. The orange curve fixes the right singular vectors at $\boldsymbol V_{\mathrm p}$ and optimizes only the IRS phase-shift matrix; the green curve jointly optimizes $\boldsymbol V$ and the IRS phase-shift matrix. The horizontal axis is $\|\boldsymbol H_{\mathrm p}\|_{\mathrm F}^{2}/[\sigma_1(\boldsymbol F)\sigma_1(\boldsymbol G)]^2$. Each marker gives the arithmetic mean of the linear normalized MSE, converted to dB. The green curve is lower at every tested energy. At normalized energies $0.0864$ and $2.16$, the mean MSE decreases from $0.435$ to $1.45\times10^{-3}$ and from $0.564$ to $0.108$, respectively. With the same propagation matrices $\boldsymbol F$ and $\boldsymbol G$, these results illustrate the benefit of choosing the channel's right singular vectors jointly with the IRS phase-shifts while retaining its prescribed left singular vectors and singular values, as allowed in \eqref{eq:isospectral_channel_customization}.

As the target energy increases, all four prescribed singular values increase with fixed ratios. A matrix-product singular-value inequality bounds the fourth generated singular value by the smallest of the four products pairing the first through fourth singular values of $\boldsymbol F$ with the fourth through first singular values of $\boldsymbol G$, respectively \cite[Corollary~3.3]{zhao2024mimo}. This bound is independent of the IRS phase-shifts, which preserve the singular values of each propagation matrix. At the highest normalized target energy of $2.16$, the fourth prescribed singular value exceeds this bound for 271 of the 300 targets. Those targets cannot attain zero customization error even after $\boldsymbol V$ is rotated. Thus, BS-side rotation reduces the error caused by unsuitable prescribed right singular vectors but cannot overcome the restrictions imposed by the propagation spectra.

\subsection{Orthogonal Right Singular Vectors and PCA Overlap}

\begin{figure}[!t]
		\centering
		\includegraphics[width=\columnwidth,trim=0 0.00pt 0 19.00pt,clip]{\detokenize{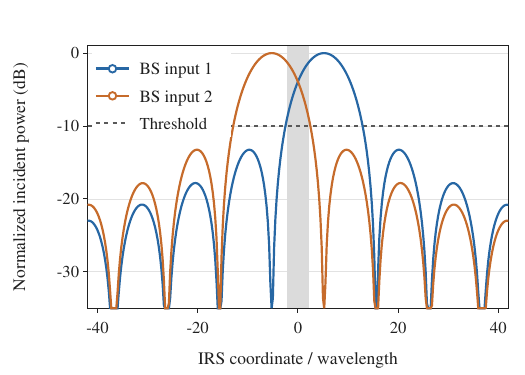}}
		\caption{PCA overlap (shaded) for orthogonal BS-side right singular vectors.}
		\label{fig:numerical_pca_overlap}
	\end{figure}

\begin{figure}[!t]
		\centering
		\includegraphics[width=\columnwidth,trim=0 0.00pt 0 19.00pt,clip]{\detokenize{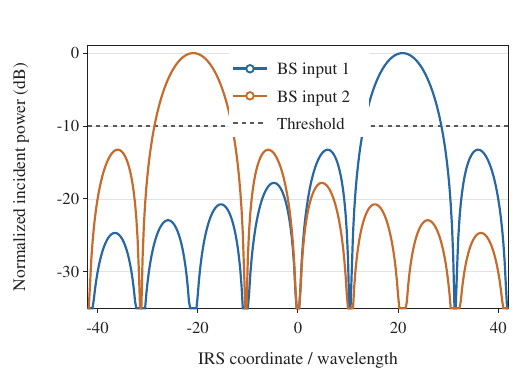}}
		\caption{Disjoint PCAs of orthogonal BS-side right singular vectors.}
		\label{fig:numerical_pca_disjoint}
	\end{figure}

Figs.~\ref{fig:numerical_pca_overlap} and \ref{fig:numerical_pca_disjoint} compare two pairs of orthogonal right singular vectors for facing arrays with $M=64$, $N=1025$, and $r_{\mathrm{BI}}=10\,\mathrm{m}$. The BS--IRS channel uses spherical-wave propagation, with each path amplitude equal to the center-to-center distance divided by the path length and by $\sqrt N$. Each unit-norm transmit vector compensates the spherical phase toward the IRS center and adds a linear phase progression across BS antennas. The two figures use the spatial-frequency index pairs $(-1/2,1/2)$ and $(-2,2)$, respectively, where one index unit gives a phase increment of $2\pi/M$ between adjacent antennas. The common spherical phase cancels in the inner products; the integer index differences of 1 and 4 ensure orthogonality in the respective pairs.

Assume that all $|\alpha_{\ell,n}|$ equal $\alpha_0$. Each curve shows incident power normalized by its peak, in decibels. The $-10$ dB threshold selects powers at least $0.1$ of the peak, or contribution amplitudes at least $\alpha_0\sqrt{0.1}$ times the peak incident amplitude. Equal peak powers within each figure make this threshold common to both streams. The shaded region in Fig.~\ref{fig:numerical_pca_overlap} is their PCA intersection, whereas Fig.~\ref{fig:numerical_pca_disjoint} has disjoint PCAs. Thus, orthogonality alone does not ensure PCA separation.

For the disjoint pair, the numerical inner-product magnitude is $1.00\times10^{-16}$. We set $\boldsymbol U_{\mathrm p}=\boldsymbol I_2$ and use far-field IRS--UE channels with direction sines $-0.20$ and $0.25$ and common element magnitude $N^{-1/2}$, yielding $\alpha_0=N^{-1/2}$. For each of 20 targets, the two prescribed singular values equal one value drawn uniformly between $20\%$ and $30\%$ of the smaller single-stream maximum in \eqref{eq:integrated_maximum}. The prescribed gains and left singular vectors then remain fixed. We minimize the full channel error by updating each IRS phase-shift to its coordinate-wise minimizer, using one zero-phase and two uniformly random initializations. Each sweep visits the IRS elements in a random order, with random seed $2026100807$. We allow at most 220 sweeps and stop when the largest decrease in squared Frobenius error over the three runs is below $10^{-10}$, retaining the lowest-error result.

With identity prescribed left singular vectors, zero error requires orthogonal generated channel rows along the selected right singular vectors. Their normalized correlation is the inner-product magnitude divided by the product of their norms. Across the 20 targets, its median is $6.85\times10^{-6}$, and the median normalized MSE is $7.31\times10^{-6}$.

\subsection{Contribution Distributions after Optimization}

To examine equal-gain customization with selectable right singular vectors, we maximize the common gain $\xi$ for three UEs under $\boldsymbol H_{\mathrm c}\boldsymbol H_{\mathrm c}^{\mathrm H}=\xi^2\boldsymbol I_3$, using 100 uniformly random phase initializations without PCA constraints. Here, $\lambda=3.75$ mm (80 GHz), $M=128$, and $N=513$. The $x$-directed BS and IRS arrays are centered at $(0,4,3)$ m and $(0,0,3)$ m; the UEs are at $(-180,300,1.5)$ m, $(-85,300,1.5)$ m, and $(180,300,1.5)$ m. Direct paths are blocked. The BS--IRS channel uses spherical waves with magnitude $4/(\sqrt{257}d_{n,m})$; the IRS--UE channel uses plane waves with magnitude $300/(\sqrt{257}r_\ell)$, where $d_{n,m}$ and $r_\ell$ denote the BS-antenna--IRS-element and IRS-center--UE distances in meters, respectively. All UE distances exceed the $122.88$ m far-field boundary. We solve the phase optimization by an augmented-Lagrangian method with limited-memory Broyden--Fletcher--Goldfarb--Shanno updates, allowing at most 20 outer iterations and 900 inner iterations, followed by a Gauss--Newton correction of the Gram constraints.

To assess the unknown optimum, we form a semidefinite relaxation of the unit-modulus phase optimization. The objective is the average squared norm of the three generated channel rows. Two row-power equalities and the real and imaginary parts of the three pairwise row inner products impose eight real quadratic equality constraints. Dropping the rank-one constraint on the lifted phase matrix gives an upper bound; a feasible point of its dual then bounds every feasible common squared gain. We verify dual feasibility by checking that the Hermitian slack matrix is positive semidefinite. Its computed smallest eigenvalue is approximately $10^{-9}$, and conservative rounding gives the squared-gain upper bound $17.08109$.

The best feasible gain is $4.07858$, while the bound gives $\xi\leq4.13293$. The relative loss of the best feasible common amplitude from the unknown global optimum is therefore at most $1-4.07858/4.13293\approx1.315\%$. The normalized Gram residual $\|\boldsymbol H_{\mathrm c}\boldsymbol H_{\mathrm c}^{\mathrm H}-\xi^2\boldsymbol I_3\|_{\mathrm F}/\|\boldsymbol H_{\mathrm c}\boldsymbol H_{\mathrm c}^{\mathrm H}\|_{\mathrm F}$ stays below $8.0\times10^{-16}$ in all 100 runs.

\begin{figure}[!t]
\centering
\includegraphics[width=\columnwidth]{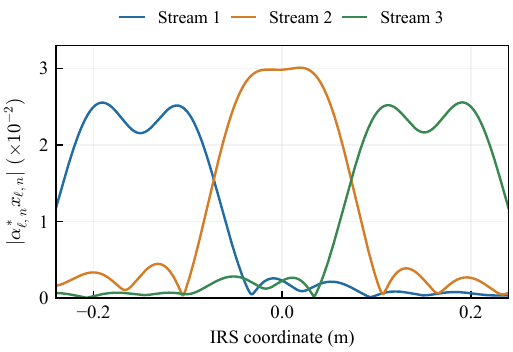}
\caption{Contribution magnitudes of the optimized three-stream channel.}
\label{fig:numerical_optimized_pca}
\end{figure}

For $\xi>0$, the Gram constraint gives $\boldsymbol V_{\mathrm s}^{\mathrm H}\boldsymbol V_{\mathrm s}=\boldsymbol I_3$ and $\boldsymbol H_{\mathrm c}\boldsymbol V_{\mathrm s}=\xi\boldsymbol I_3$, where $\boldsymbol V_{\mathrm s}=\boldsymbol H_{\mathrm c}^{\mathrm H}/\xi$. Thus, with prescribed $\boldsymbol U_{\mathrm p}=\boldsymbol I_3$, the columns of $\boldsymbol V_{\mathrm s}$ are right singular vectors for three equal singular values $\xi$. We recover $\boldsymbol v_\ell=\boldsymbol H_{\mathrm c}^{\mathrm H}\boldsymbol e_\ell/\xi$, where $\boldsymbol e_\ell$ is column $\ell$ of $\boldsymbol I_3$. Fig.~\ref{fig:numerical_optimized_pca} plots $|\alpha_{\ell,n}^{*}x_{\ell,n}|$ for the best feasible IRS phase-shift matrix and the recovered right singular vectors. For each stream, the largest values occur on a different section of the IRS, while nonzero values remain outside that section. These observations concern only the reported feasible solution.\looseness=-1

\subsection{BS--IRS Singular-Value Distribution}

\begin{figure}[!t]
		\centering
		\includegraphics[width=\columnwidth,trim=0 3.00pt 0 11.50pt,clip]{\detokenize{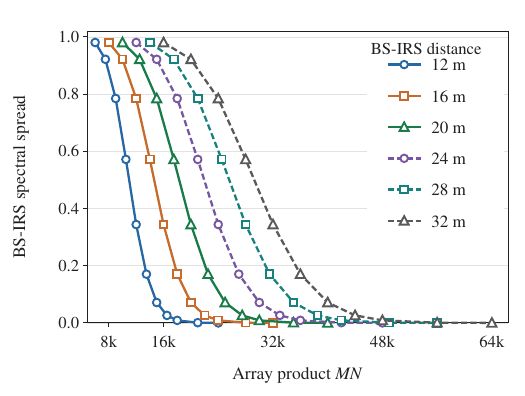}}
		\caption{BS--IRS spectral spread $1-\sigma_4^2(\boldsymbol F)/\sigma_1^2(\boldsymbol F)$ versus $MN$.}
		\label{fig:numerical_bsirs_spectral_spread}
	\end{figure}

Fig.~\ref{fig:numerical_bsirs_spectral_spread} examines the normalized Fresnel BS--IRS channel for four streams, with $M=128$, varying $N$, and six link distances. Its ordinate, $1-[\sigma_4(\boldsymbol F)/\sigma_1(\boldsymbol F)]^2$, approaches zero as the first four singular values approach equality. At every tested distance, it decreases with $MN$; at fixed $MN$, it increases with $r_{\mathrm{BI}}$. The parameter $c_{\mathrm{BI}}=MN d_{\mathrm B}d_{\mathrm I}/(\lambda r_{\mathrm{BI}})$ in \eqref{eq:necessary_13} combines these variables. At sampled points with $c_{\mathrm{BI}}=3.84$, the six plotted values differ by at most $7.66\times10^{-5}$. Thus, over the tested range with fixed $M$, longer links require larger $N$ for the same ratio. Equations \eqref{eq:necessary_2} and \eqref{eq:necessary_2_G} constrain the fourth singular values of both propagation matrices. This figure evaluates only $\boldsymbol F$; the spectrum of $\boldsymbol G$, prescribed receive directions, and a common IRS phase-shift matrix must also be checked before drawing conclusions about MSE or \mbox{channel customization feasibility.}\looseness=-1

\section{Conclusion}
	\label{sec:conclusion}

We studied phase-only IRS channel customization under three criteria that progressively relax the required right singular vectors and common gain scale. The equivalent counterpart problem exposes the phase constraints shared by all streams. Subchannel-gain bounds and PCA-overlap analysis guide a Fresnel construction with separately chosen focal spacing and allocation width. Its sufficient conditions retain all reflected components and require common-phase matching and nulling. Spectral necessary conditions exclude targets unsupported by the propagation channels at the specified gain scale and tolerance. The sufficient aperture condition and necessary size restriction exhibit an array-size-product dependence, although the product alone does not determine feasibility. Numerical results show that orthogonal right singular vectors need not separate PCAs and that joint vector and phase-shift selection reduces MSE in the tested settings. Array selection must therefore account for the prescribed channel, both propagation matrices, and the available phase control.

\appendices
\section{Proof of Proposition~\ref{prop:cascaded_spectral_upper_bound}}
\label{app:necessary_array_bound}

By \eqref{eq:necessary_5}, it suffices to bound $\|\boldsymbol F\|_{\mathrm F}^2/\sigma_1^2(\boldsymbol F)$ above in terms of the array dimensions. Expanding the quadratic phase in \eqref{eq:integrated_channel} separates the terms containing $q_n^2$, $p_m^2$, and $q_np_m$. The first two produce row and column phase factors, respectively. Removing them by diagonal unitary multiplication, along with the common propagation phase, preserves the singular values. After dividing by the common nonzero coefficient $\beta$, denote the remaining matrix by $\widetilde{\boldsymbol F}$, whose entries are $[\widetilde{\boldsymbol F}]_{n,m}=\exp(\jmath 2\pi q_np_m/(\lambda r_{\mathrm{BI}}))$. Since $\|\boldsymbol F\|_{\mathrm F}^2=|\beta|^2MN$ and $\sigma_1^2(\boldsymbol F)=|\beta|^2\sigma_1^2(\widetilde{\boldsymbol F})$, the common coefficient cancels, \mbox{and we obtain}
\begin{equation}
\frac{\|\boldsymbol F\|_{\mathrm F}^2}{\sigma_1^2(\boldsymbol F)}=\frac{MN}{\sigma_1^2(\widetilde{\boldsymbol F})}.
\label{eq:necessary_normalized_ratio}
\end{equation}

It remains to bound $\sigma_1(\widetilde{\boldsymbol F})$ below. For $0<c_{\mathrm{BI}}\le1/2$, the centered array coordinates give $|q_n|\le(N-1)d_{\mathrm I}/2$ and $|p_m|\le(M-1)d_{\mathrm B}/2$. Hence, each entry phase has magnitude at most $\pi c_{\mathrm{BI}}/2\le\pi/4$, and each real part is at least $\cos(\pi c_{\mathrm{BI}}/2)$. To use this entrywise bound, multiply $\widetilde{\boldsymbol F}$ on the left by $N^{-1/2}[1,\ldots,1]$ and on the right by $M^{-1/2}[1,\ldots,1]^{\mathrm T}$. These unit-norm vectors give the normalized sum of the entries, whose magnitude is bounded above by the largest singular value. Therefore,
\begin{gather}
\sigma_1(\widetilde{\boldsymbol F})\overset{(a)}{\ge}\frac{\left|\sum\limits_{n=1}^{N}\sum\limits_{m=1}^{M}[\widetilde{\boldsymbol F}]_{n,m}\right|}{\sqrt{MN}}\overset{(b)}{\ge}\sqrt{MN}\cos\!\left(\frac{\pi c_{\mathrm{BI}}}{2}\right),
\label{eq:necessary_8}\\
\frac{MN}{\sigma_1^2(\widetilde{\boldsymbol F})}\le\frac{1}{\cos^2(\pi c_{\mathrm{BI}}/2)}\overset{(c)}{\le}1+\frac{\pi^2c_{\mathrm{BI}}^2}{2}\overset{(d)}{\le}1+8c_{\mathrm{BI}}.
\label{eq:necessary_small_c_ratio}
\end{gather}
Inequality (a) uses the fact that both multiplying vectors have unit norm, and (b) sums the entry real parts. For (c), write $\cos^{-2}(\pi c_{\mathrm{BI}}/2)=1+\sin^2(\pi c_{\mathrm{BI}}/2)/\cos^2(\pi c_{\mathrm{BI}}/2)$ and use $\sin(\pi c_{\mathrm{BI}}/2)\le\pi c_{\mathrm{BI}}/2$ and $\cos^2(\pi c_{\mathrm{BI}}/2)\ge1/2$. Inequality (d) uses $c_{\mathrm{BI}}\le1/2$ and $\pi^2/4<8$.

For $1/2<c_{\mathrm{BI}}\le MN/2$, the phases of the full matrix need not lie within the range used in \eqref{eq:necessary_8}. We instead select $n_0$ consecutive rows and $m_0$ consecutive columns, with $1\le n_0\le N$ and $1\le m_0\le M$, so that the same argument can be applied to a submatrix. Recenter the selected coordinates at their respective midpoints. Expanding $q_np_m$ then gives the product of the centered offsets, a row-dependent term, a column-dependent term, and a constant. Multiplication by diagonal unitary matrices removes the latter three terms without changing the singular values. Denote the resulting matrix by $\widetilde{\boldsymbol F}_{\mathrm{sub}}$. The centered offsets have magnitudes at most $(n_0-1)d_{\mathrm I}/2$ and $(m_0-1)d_{\mathrm B}/2$, respectively, so each remaining entry phase has magnitude at most
\begin{equation}
\frac{\pi c_{\mathrm{BI}}(n_0-1)(m_0-1)}{2MN}
\le\frac{\pi c_{\mathrm{BI}}n_0m_0}{2MN}.
\label{eq:necessary_9}
\end{equation}

To retain a real part of at least $1/\sqrt2$ in every entry, it suffices to make the bound in \eqref{eq:necessary_9} no larger than $\pi/4$, or $n_0m_0\le MN/(2c_{\mathrm{BI}})$. Set $T=MN/(2c_{\mathrm{BI}})$ and choose
\begin{equation}
(n_0,m_0)=
\begin{cases}
(\lfloor T\rfloor,1),&N\ge T,\\
(N,\lfloor T/N\rfloor),&N<T .
\end{cases}
\label{eq:necessary_10}
\end{equation}
Here, $1\le T<MN$. If $N\ge T$, then $1\le\lfloor T\rfloor\le N$, while $m_0=1\le M$. If $N<T$, then $1<T/N<M$, so $1\le\lfloor T/N\rfloor<M$ and $n_0=N$. Thus, both choices give admissible submatrix dimensions. Moreover, $\lfloor x\rfloor\ge x/2$ for $x\ge1$ gives $T/2\le n_0m_0\le T$. The submatrix therefore contains at least $T/2$ entries while meeting the required phase bound.

Using unit vectors $n_0^{-1/2}[1,\ldots,1]$ and $m_0^{-1/2}[1,\ldots,1]^{\mathrm T}$ gives a normalized entry sum with real part at least $\sqrt{n_0m_0/2}$. Its magnitude is no greater than the submatrix's spectral norm. Restricting rows and columns cannot increase the spectral norm, so \mbox{the full matrix satisfies}
\begin{equation}
\sigma_1^2(\widetilde{\boldsymbol F}) \overset{(a)}{\ge}\sigma_1^2(\widetilde{\boldsymbol F}_{\mathrm{sub}}) \overset{(b)}{\ge}\frac{n_0m_0}{2}\ge\frac{T}{4} =\frac{MN}{8c_{\mathrm{BI}}}.
\label{eq:necessary_11}
\end{equation}
For (a), the row and column phase removal preserves the submatrix norm, and its unit vectors can be extended by zeros to the full dimensions. Inequality (b) uses the entry sum just obtained. The choice in \eqref{eq:necessary_10} ensures $n_0m_0\ge T/2$, yielding $MN/\sigma_1^2(\widetilde{\boldsymbol F})\le8c_{\mathrm{BI}}$. Selecting a submatrix only estimates $\sigma_1(\widetilde{\boldsymbol F})$; all IRS elements remain in \mbox{the original channel}.

For $c_{\mathrm{BI}}>MN/2$, selecting a single entry of magnitude one with unit coordinate vectors gives $\sigma_1(\widetilde{\boldsymbol F})\ge1$. Hence,
\begin{equation}
\frac{MN}{\sigma_1^2(\widetilde{\boldsymbol F})}
\le MN<2c_{\mathrm{BI}} .
\label{eq:necessary_12}
\end{equation}

These cases establish $MN/\sigma_1^2(\widetilde{\boldsymbol F})\le1+8c_{\mathrm{BI}}$ for every $c_{\mathrm{BI}}>0$. Substitution into \eqref{eq:necessary_normalized_ratio} gives \mbox{the following bound.}
\begin{equation}
\frac{\|\boldsymbol F\|_{\mathrm F}^2}{\sigma_1^2(\boldsymbol F)}
\le1+8c_{\mathrm{BI}} .
\label{eq:necessary_13}
\end{equation}
Combining this upper bound with \eqref{eq:necessary_5}, subtracting $1$, and substituting $c_{\mathrm{BI}}=MN d_{\mathrm B}d_{\mathrm I}/(\lambda r_{\mathrm{BI}})$ yields \eqref{eq:necessary_6}, \mbox{proving the proposition}.\hspace*{\fill}\nobreakspace\IEEEQED

\bibliographystyle{IEEEtran}
	\bibliography{references}
	
\end{document}